\documentclass{amsart}
\usepackage{amssymb,stmaryrd}
\usepackage{amsfonts}
\usepackage{amstext}
\usepackage{geometry}                
\usepackage{graphicx}    
\usepackage{amssymb,amsmath,epsfig}
\usepackage{scalefnt}
\usepackage{MnSymbol}
\usepackage[all]{xypic}
\usepackage{slashed}
\usepackage{extarrows}
\usepackage{stmaryrd}
\usepackage{lettrine}
\usepackage{booktabs}
\usepackage{paralist}
\usepackage{subfig}
\newtheorem{lemma}{Lemma}[section] 
\newtheorem{proposition}[lemma]{Proposition}

\renewcommand{\imath}{\mathrm{i}}

\newcommand*{\doublenabla}{%
  \nabla\mkern-12mu\nabla
}   

\newcommand{\C}{\mathbb{C}}
\newcommand{\R}{\mathbb{R}}

\newcommand{\Z}{\mathbb{Z}}

\newcommand{\X}{v_x}
\newcommand{\Y}{v_t}

\newcommand{\del}{\partial}

\newcommand{\ev}{\mathrm{ev}}

\newcommand{\extd}{\mathrm{d}}

\newcommand{\tens}{\mathop{{\otimes}}}

\newcommand{\id}{\mathrm{id}}

\newcommand{\<}{\langle}
\renewcommand{\>}{\rangle}

\newcommand{\Tr}{\mathrm{ Tr}}
\renewcommand{\div}{\mathrm{div}}

\allowdisplaybreaks

\begin{document}

\author{Chengcheng Liu and Shahn Majid}
\address{School of Mathematical Sciences\\ Queen Mary University of London \\ Mile End Rd, London E1 4NS }
\email{ chengcheng.liu@qmul.ac.uk, s.majid@qmul.ac.uk}
\thanks{Ver.1 The first author was supported by a China Scholarship Grant}

\title{Quantum geodesics on $\lambda$-Minkowski spacetime}
	\begin{abstract} We apply a recent formalism of quantum geodesics to the well-known bicrossproduct model $\lambda$-Minkowski quantum spacetime $[x^i,t]=\imath\lambda_p x^i$ with its flat quantum metric as a model of quantum gravity effects, with $\lambda_p$ the Planck scale. As examples, quantum geodesic flow of a plane wave gets an order $\lambda_p$ frequency dependent correction to the classical geodesic velocity. A quantum geodesic flow with classical velocity $v$ of a Gaussian with width  $\sqrt{2\beta}$ initially centred at the origin changes its shape but its centre of mass moves with ${\<x\>\over\<t\>}=v(1+{3\lambda_p^2\over 2\beta}+O(\lambda^3_p))$, an order $\lambda_p^2$ correction. This implies, at least within perturbation theory, that a `point particle' cannot be modelled as an infinitely sharp Gaussian due to quantum gravity corrections.  For contrast, we also look at quantum geodesics on the noncommutative torus with a 2D curved weak quantum Levi-Civita connection.  
	\end{abstract}
\maketitle
\section{Introduction}

The {\em quantum spacetime hypothesis}, that spacetime could be better modelled by noncommutative coordinates due to quantum gravity effects, has gained increasing interest in recent years, while also a source of speculation since the arrival of quantum mechanics. An early model was in \cite{Sny} although not with a closed spacetime algebra as such. First models of quantum gravity unified with gravity with noncommutative but curved (phase) space appeared in \cite{Ma:pla}. The interpretation of flat but noncommutative spacetime as classical but curved momentum space appeared in \cite{Ma:dua}, a family which includes the much-studied bicrossproduct model $\lambda$-Minkowski spacetime\cite{MaRue} 
\[ [x^i,t]=\imath\lambda_p x^i\]
with quantum Poincar\'e group isomorphic to a proposal in \cite{Luk} and with potentially testable but heuristic predictions\cite{AmeMa}. This is also called $\kappa$-Minkowski spacetime with $\kappa=1/\lambda_p$, but we will work with  $\lambda_p$ which, in the context of quantum gravity corrections, could be expected to be of order the Planck scale. More precisely, we will work with $\lambda=\imath\lambda_p$. Other early flat quantum spacetime proposals included \cite{Hoo,DFR}. Since these early days, a systematic quantum Riemannian geometry over a noncommutative algebra $A$ has emerged as in the recent text\cite{BegMa} and references therein, including a theory of metrics $g\in \Omega^1\tens_A\Omega^1$, quantum Levi-Civita connections $\Omega^1\to \Omega^1\tens_A\Omega^1$ on 1-forms $\Omega^1$, and curvature. Using this formalism, the first models of quantum gravity itself on finite quantum spacetimes have emerged\cite{Ma:sq,LirMa,ArgMa1,ArgMa2}, giving a glimpse of what this entails as visible from such baby models. Effects include a uniform level of vacuum fluctuations of the metric and remarkable reality properties in the deep quantum gravity regime. Particle creation in curved noncommutative backgrounds have also been studied in this approach\cite{Ma:haw,ArgMa1} but there is much that remains to be studied even on the currently known models. 

In the present paper, we come full circle and apply this new geometric and coordinate-invariant formalism back to the flat $\lambda$-Minkowski spacetime. 30 years later, we no longer need to make heuristic and ad-hoc interpretations but can just see what the theory says about issues such as the effective speed of light\cite{AmeMa} in this model. The formalism requires to first fix the differential structure by specifying $\Omega^1$, and here it has been known for a long time that there isn't  one of the expected dimension that is quantum-Poincar\'e invariant, but rather there is an extra non-classical 1-form $\theta'$\cite{Sit} needed to maintain the quantum-Poincar\'e invariance. We use this standard calculus. This is an example of a rather common phenomenon in quantum spacetime model building, a kind of quantum anomaly for differential structures resolved either by nonassociativity or by adding an extra dimension. The first new ingredient for the quantum Riemannian geometry is the quantum metric and we just use the quantum Poincare invariant one and its quantum Levi-Civita connection\cite[Prop.~9.4]{BegMa}
\[ g=-\extd t\tens\extd t+\sum_i \extd x^i\tens\extd x^i + \theta\tens\theta,\quad \nabla(\extd x^i)=\nabla(\extd t)=\nabla\theta=0\] where $\theta=\theta'+\extd t$ is the more convenient extra direction here. There are other natural quantum differential structures on this algebra without the extra dimension but these only admit curved quantum geometries\cite{BegMa:gra,MaTao:cos} whereas we want to focus on flat $\lambda$-Minkowski spacetime. We will  briefly return to curvature at the end in Section~\ref{sectorus}, but for the much simpler model of the well-known noncommutative torus with its standard 2D $\Omega^1$ and now with its moduli of curved (weak) quantum Riemannian geometries recently found in \cite{BegMa}. 

Next, to explore particle motion, we will use the recently introduced formalism of `quantum geodesics' \cite{Beg:geo,BegMa:geo, BegMa:ric} in quantum Riemannian geometry.  The mathematical formalism is recalled in  the Preliminaries Section~\ref{secpre} and centres on the notion of an $A$-$B$-bimodule connection $\nabla_E:E\to E\tens_B\Omega^1_B$, see \cite{BegMa}, where $E$ is an $A$-$B$-bimodule.  Here $A$ is the quantum spacetime coordinate algebra of interest and $B$ is the geodesic time coordinate algebra which in our case we will keep classical, $B=C^\infty(\R)$. The construction also depends on the choice of geometric connection $\nabla$ on $\Omega^1$ of $A$, but not directly on the quantum metric itself.  As the formalism in \cite{Beg:geo,BegMa:geo, BegMa:ric} is rather more abstract than one might wish, Section~\ref{secpara} analyses how everything looks in `local coordinates' or more precisely when we have a basis of 1-forms. This offers more explicit  formulae, which will be useful for computations in other models also. 

Without going into any of the mathematics yet, we first explain the ideas. In fact,  quantum geodesics entail a new way of thinking about geodesics even on classical Euclidean space or Minkowski spacetime. Full details will be given in Section~\ref{secflat} but the first idea is to think not of a single particle geodesic (which of course is just a straight line) but of a `fluid' of particles where each one moves along a geodesic. If, initially, these are moving in various directions with various speeds then after a while an initially uniform distribution will become non-uniform. In fact we describe this  quantum mechanically  by an `amplitude' half density $\psi$ where $|\psi|^2$ would be the `fluid' density or more precisely a probability density. In the classical case a real $\psi$ will remain real and whether we work with the density or $\psi$ does not make much difference. But the first thing we will see in the $\lambda$-Minkowski model is $O(\lambda_p)$ complex corrections to it, which can then potentially interfere and diffract in analogy with quantum mechanics but with respect to geodesic time.  Also, returning to our `fluid' picture, the tangent vectors of all our particles moving in different directions form a vector field $X$ which will also evolve in geodesic time. Beggs' discovery in \cite{Beg:geo} is that the evolution of this velocity field $X$ can be intrinsically characterised by a certain {\em geodesic velocity equation}.  Thus, one can first solve for  $X$ and then afterward the amplitudes $\psi$ obey a Schroedinger-like {\em amplitude flow equation} with respect to $X$ that expresses its role as the tangents to the particle flow. In effect, we rip apart the concept of a geodesic into two concepts, a geodesic velocity flow and then the geodesic amplitude flow  relative to it. All of this takes some getting used to and Section~\ref{secflat} gives some in-depth examples in this flat classical case. On the other hand, this point of view makes sense in quantum geometry without mention of individual points or curves. One can also show\cite{BegMa:geo} that the actual Schroedinger equation can be viewed as a quantum geodesic amplitude flow on the Heisenberg algebra. 

When $A$ is thought of as functions on space then the geodesic time parameter external to that would be time, and this is our fluid or wave function picture. The picture is less clear and only exists by analogy with that when $A$ is the algebra of functions on {\em spacetime}. In this case, we will refer to the geodesic time parameter as $s$ (to avoid confusion with the time coordinate of spacetime), and any notion of probability in the interpretation of $\psi$ is with respect to this and not with respect to the spacetime time.  For ordinary geodesic motion in GR, for a single massive particle,  $s$ has an interpretation as the proper time experienced along the geodesic, but what happens when there are many particles? To address this conundrum we can imagine that $\psi$ at $s=0$ is localised around the origin in spacetime and then watch how it evolves. Not surprisingly, we find for classical flat spacetime that the bump moves in a straight line at a constant speed, even though $X$ may be far from constant. We prove this in detail in Proposition~\ref{psisol} for the simpler case of $X$ as most linear in the spacetime cordinates, where we solve the geodesic flow explicitly. Indeed, the solutions are controlled by a single diffeomorphism with inverse $\phi_s^{-1}(x^\mu)=x^\mu-s X_0^\mu(x)$, where $X_0$ is the initial value of the velocity field as $s=0$. This then gives an `emergent proper time' interpretation of $s$ that applies to the evolution of all initial amplitudes. These may expand or contract to reflect changing density by a factor in the solution that depends on $\kappa={1\over 2}{\rm div}(X)$. We also analyse when flow is time-like. 

Section~\ref{secbic} now turns the the $\lambda$-Minkowski quantum spacetime case, starting with the case in Section~\ref{secon} where the components of $X$ are constant on spacetime (they are then forced to also be constant in geodesic time). Classically, the whole spacetime in this case flows with a constant velocity $v$. We show how to solve the amplitude flow equation in general using perturbation theory and apply it to initial $\psi$ as $s=0$ given by a Gaussian of  `width' $\sqrt{2\beta}$ (or twice the standard deviation for a normal distribution). As $s$ increases, the Gaussian does not retain its shape but the expectation values $\<x\>$ and $\<t\>$ behave simply and  the effective velocity gets a leading correction governed by ${3\lambda^2_p\over 2\beta}$. This makes sense on dimensional grounds but is only valid for Gaussian width well above the Planck scale. This has the startling consequence that one cannot physically think of a point-particle as an infinitely sharp Gaussian because of quantum gravity effects. 

We also solve the amplitude flow equation exactly, without perturbation theory, for initial $\psi$ a plane wave of frequency $\omega$. In this case we find that effect of the quantum spacetime is an order $\lambda_p\omega$ correction to the effective geodesic velocity $v$. This is reminiscent of early claims in \cite{AmeMa} about a frequency-dependent variable speed of light but now given an operational meaning which was missing before. In our analysis, it is not the energy or momentum components of the plane wave which are being modified but the phase shift due to the constant geodesic motion. 

 In Section~\ref{seclin} we similarly look at quantum geodesic velocity fields $X$ that are at most linear in the spacetime coordinates. There are fewer solutions than classically, i.e., not all classical geodesic flows quantise in the strict formulation of \cite{Beg:geo, BegMa:geo}, but we find several families. Some of these both classically and at the quantum level develop singularities at finite $s=a$, say, and here we find that the order $\lambda$ quantum corrections to the associated $\psi$ blow up faster as $s\to a$, i.e. the perturbation theory is not valid within $\lambda_p$ from $a$. Section~\ref{secrem} concludes with some discussion of these matters and directions for further work. We assume units in which the usual speed of light $c=1$, or more precisely where $x,t$ are considered to have the same units.

\section{Preliminaries} \label{secpre}

Here we recall what is a quantum geodesic as proposed in\cite{Beg:geo} and studied further in \cite{BegMa:geo,BegMa:ric}. We explain the algebraic point of view which works even when the `coordinate algebra' of the spacetime is a possibly noncommutative unital algebra $A$. 

\
\subsection{Elements of quantum Riemannian geometry formalism} 

We very briefly outline the context of quantum Riemannian geometry\cite{BegMa}, although we will not need the full power of the formalism as most of the paper concerns flat quantum spacetime where the $\nabla=0$ in a suitable basis. The formalism is very different in character from the well-known Connes approach\cite{Connes}, but not incompatible with it, growing instead out of model building with quantum group symmetry\cite{BegMa:rie} and quantum spacetime\cite{BegMa:gra}. 

If $A$ is the coordinate algebra, we first need to fix a `differential calculus' in the sense of an exterior algebra $(\Omega,\extd)$ with $A$ appearing in degree 0 and $\extd$ squaring to zero and obeying the graded-Leibniz rule. In degree 1 it mean a space $\Omega^1$ which is a bimodule in the sense that one can multiply by $A$ from the left and the right and the two actions commute, and $\extd:A\to \Omega^1$ needs to obey the Leibniz rule 
\[ \extd(ab)=a\extd b+ (\extd a)b\]
for all $a,b\in A$. We also ask that $\Omega^1$ is spanned by elements of the form $a\extd b$ for $a,b\in A$. Classically, $\Omega^1$ would be elements of the form $f_\mu\extd x^\mu$ in local coordinates. We work over $\C$ and we suppose a $*$-involution on the algebra, extending to forms so as to commute with $\extd$. This is needed  to recover classical real geometry in the classical case with self-adjoint generators.

 A quantum metric is then $g\in\Omega^1\tens_A\Omega^1$ which is invertible in the sense of a bimodule map $(\ ,\ ):\Omega^1\tens_A\Omega^1$ obeying the usual requirements as inverse to $g$. This forces $g$ in fact to be central. In GR one writes $g=g_{\mu\nu}\extd x^\mu\extd x^\nu$  but there is an implicit tensor product over the algebra of functions here. For linear connections, we similarly work with $\nabla:\Omega^1\to\Omega^1\tens_A\Omega^1$ which classically would appear as $\nabla\extd x^\mu=-\Gamma^\mu_{\nu\rho}\extd x^\nu\tens\extd x^\rho$ and where classically we would evaluate the left factor against a vector field to get the covariant derivative along a vector field and therefore not write any tensor product. Here we will work more abstractly and also use conventions where we evaluate vector fields on the right factor. Such a {\em right connection} is characterised by
\[ \nabla(\omega.a)=\omega\tens\extd a+ (\nabla\omega).a,\quad \nabla(a.\omega)=\sigma(\extd a\tens\omega)+a\nabla\omega\]
for all $a\in A, \omega\in \Omega^1$. Here, the `generalised braiding' $\sigma:\Omega^1\tens_A\Omega^1\to \Omega^1\tens_A\Omega^1$ is assumed to exist and is uniquely determined by the second equation. A quantum Levi-Civita connection (QLC) means such a bimodule connection which is metric compatible and torsion free in the sense
\[ \nabla g:=((\id\tens\sigma)(\nabla\tens\id)+ \id\tens\nabla)g=0, \quad \wedge\nabla+\extd=0\]
Sometimes a QLC does not exist and we may be interested in the larger moduli of weaker WQLCs.  There are also `reality' properties needed for both $g$ and $\nabla$ with respect to $*$. More details are in \cite{BegMa}. The notion of a `bimodule connection' used here goes back to \cite{DVM,Mou}. 

\subsection{Elements of quantum geodesic formalism} 

We start with a very brief outline of the abstract quantum geodesic theory based on a pair of algebras $A,B$ and an $A-B$-bimodule $E$ (where one can multiply from the left by $A$ and from the right by $B$ and the two actions commute). Details are in \cite{BegMa:geo,BegMa:ric} and again our goal in the paper is to make this all very concrete by looking at an simple example. The theory is general but we will be concerned with $E$ only form $E=A\tens B$, where $B=C^\infty(\R)$ is the `geodesic time coordinate algebra' and $A$ is the quantum spacetime algebra where the geodesic flow takes place. Clearly, an element $\psi\in E$ then just means a time-dependent element $\psi_t\in A$ for each geodesic time $t$. Both $A$ and $B$ need differential structures $\Omega^1,\Omega^1_B$ respectively but the latter will be the classical one with basis $\extd t$, and these both need (right) linear connections but the one on $\Omega^1_B$ will be trivial $\nabla_B\extd t=0$. 

We first thing to note that the above notion of a right bimodule connection on $\Omega^1$ as an $A$-$A$-bimodule works the same way on any  $A-B$ bimodule $E$\cite{BegMa}, with $\nabla_E: E\to E\tens_B \Omega^1_B$ now such that\cite{BegMa}
\[  \nabla_E(\psi b)=\psi\tens\extd b+(\nabla_E a)b,\quad \nabla_E(a\psi)=\sigma(\extd a\tens \psi)+a\nabla_E \psi   \]
for all $\psi\in E$, $a\in A$ and $b\in B$, for some bimodule map $\sigma_E: \Omega^1\tens_A E\to E\tens_B\Omega^1_B$. Also observe that both sides of the map $\sigma_E$ have tensor product $A$-$B$-bimodule connections given the connections on $\Omega^1,\Omega^1_B,E$, so there is a well-defined covariant derivative of $\sigma_E$ as a tensor, denoted $\doublenabla(\sigma_E)$. The vanishing of this would say that $\sigma_E$ intertwines the connections before and after the map is applied.  

In our case, we set $E=A\tens B$ as mentioned and let $X_t:\Omega^1\to A$ at each $t$ be a left vector field (meaning a left module map) and $\kappa_t\in A$ a time dependent element of $A$, then we define a particular $A$-$B$-bimodule connection by
\begin{equation}\label{nablaE}\nabla_E \psi=(\dot \psi+X_t(\extd \psi)+\psi\kappa_t)\tens\extd t,\quad \sigma_E(\omega\tens \psi)=X_t(\omega.\psi)\tens\extd t.\end{equation}
It is then explained in detail in \cite{Beg:geo,BegMa:geo} that $\doublenabla(\sigma_E)=0$ is equivalent to the {\em geodesic velocity equation} 
\begin{equation}\label{veleq} \dot X_t(\omega)+ [X_t,\kappa_t](\omega)+X_t(\extd X_t(\omega))-X_t(\id\tens X_t)\nabla\omega=0 \end{equation}
for all $\omega\in \Omega^1$
and the {\em auxiliary equation}
\begin{equation}\label{auxeqn} X_t(\id\tens X_t)(\sigma-\id)=0.\end{equation}
 Note in (\ref{veleq}) that $[X_t,\kappa_t](\omega)=X_t(\omega)\kappa_t-X_t(\omega \kappa_t)$ according to the canonical actions of $A$ on the space of left vector fields. Also note that (\ref{auxeqn}) is empty classically. We will see in our examples that it can be solved but is a little restrictive. It is proposed in recent work \cite{BegMa:ric} that one can slightly weaken $\doublenabla(\sigma_E)=0$ so as to drop (\ref{auxeqn}) and then replace it by something weaker, but here we stick with the original formulation. 

Finally, for the physical interpretation, we need a `measure' in the sense of a positive linear functional $\int: A\to \C$,  subject as explained in \cite{Beg:geo,BegMa:geo} to the {\em unitarity conditions}
 \begin{equation}\label{unitarity} \int \kappa^*_t a+a\kappa_t+X_t(\extd a)=0,\quad \int X_t(\omega^*)-X_t(\omega)^*=0\end{equation}
for all $a\in A$ and all $\omega\in\Omega^1$. 

 We think of $\psi$, as geodesic-time dependent `wave-function' element of $A$ (or more precisely $\psi$ is an element of a completion of $E$ to a Hilbert $A$-$B$-bimodule) with $\rho=\psi^*\psi$ a geodesic-time dependent probability density with respect to $\int$, and we impose the {\em amplitude flow equation} $\nabla_E\psi=0$, or explicitly
 \begin{equation}\label{ampeqn}\dot \psi=-X_t(\extd \psi)-\psi\kappa_t. \end{equation}
This equation classically would say that if $\psi$ is the amplitude for particles with probability density $\rho$ and if these were actual particles in a fluid analogy then these would be moving with velocity $X_t$ at any given geodesic time $t$\cite{Beg:geo}. This is a quantum-mechanics like picture with respect to $t$ as the external geodesic time. 

In fact, $X_t$ also accounts for the changing density and unless  $X_t=X$ independently of $t$, we will see that it is not exactly the same as the effective flow velocity as tested by bump functions in the classical case. Here, in the time-independent classical case we can exponentiate $-X$ to a 1-parameter group of diffeomorphisms
 \[ \phi_t(x);\quad \dot\phi_t(x)=-X(\phi_t(x)),\quad \phi_t\circ\phi_{t'}=\phi_{t+t'},\quad \phi_t^{-1}=\phi_{-t};\quad \dot\phi_t^{-1}(x)=X(\phi_t^{-1}(x))\] 
which can be used to solve the flow equation at least if $\div X=0$, when  $\psi_t(x)=\psi_0(\phi_t(x))$ for any initial $\psi_0$. Note that a bump function $\psi_0$ centred at $x_0$ moves to one centred at $\phi^{-1}_t(x_0)$, so it is this which is the effective flow in our terms. When $X_t$ depends on geodesic time then we can similarly use a geodesic-time path-ordered exponential (or solve a first order differential equation) as
\[ \psi_t=Pe^{-\int^t_0 (\hat X_s+ \kappa_s)\extd s}\psi_0\]
to get $\phi_t(x)$), where $\hat X_s$ is the vector field acting as an operator on functions and $\kappa_s$ also acts as on operator by multiplication. However, this is not necessarily controlled by a 1-parameter group of diffeomorphisms $\phi_t$ solving $\dot\phi_t(x)=-X_t(\phi_t(x))$ with result that $\phi_t^{-1}(x)$ has a very different rate of change from $X_t$. We will illustrate this later in Section~\ref{secflat}. Returning to the quantum formulation,

 \begin{lemma}\label{lemuni}  If   (\ref{unitarity}) holds and $\psi$ evolves according to $\nabla_E\psi=0$ then $\int\psi^*\psi$ is constant in time.
\end{lemma}
\proof This is a consequence of $\nabla_E$ preserving the hermitian inner product, but here we give a direct check for completeness. Leaving the time dependence understood, we have $\dot\psi=-X(\extd \psi)-\psi\kappa$ and hence $\dot\psi^*=-X(\extd \psi)^*-\kappa^*\psi^*$ and we have $\rho=\psi^*\psi=\rho^*$. Hence
\begin{align*} {\extd\over\extd t}\int\rho&=\int\dot\psi^*\psi+\psi^*\dot\psi=-\int X(\extd\psi)^*\psi+\kappa^*\rho+\psi^*X(\extd\psi)+\rho\kappa=\int X(\extd \rho)-(X(\extd\psi))^*\psi-\psi^*X(\extd\psi)\\
&=\int X(\extd(\psi^*\psi))-(X(\extd\psi))^*\psi-\psi^*X(\extd\psi)=\int  X(\extd \psi^*.\psi)-X(\extd\psi)^*\psi \\
&=\int  X((\psi^*\extd \psi)^*)-X(\extd\psi)^*\psi =\int  (X(\psi^*\extd \psi))^*-X(\extd\psi)^*\psi =\int  (\psi^*X(\extd \psi))^*-X(\extd\psi)^*\psi =0
\end{align*}
where we used the first part of (\ref{unitarity}) for the 3rd equality and the second part in the 7th. We also used the Leibniz rule and that $X$ is a left module map.  \endproof

The choice of $\int$ represents the base measure with respect to which we would classically interpret $\rho=\psi^*\psi$ as a probability distribution. Classically, for the Levi-Civita connection on a Riemannian manifold $M$, we would take the Riemannian integration measure (using $\sqrt{|g|}$),  which can be characterised as having the property that
\begin{equation}\label{divcond} \int a {\rm div}_\nabla(X)+X(\extd a)=0\end{equation}
for all functions $a$ and all vector fields $X$. Following \cite{BegMa:ric}, we adopt this also in the noncommutative case where ${\rm div}_\nabla(X)=\ev\circ \hat\nabla X$ is the divergence defined by a corresponding left connection $\hat\nabla$ on the space of left vector fields followed by evaluation $\ev$ of a left vector field on a 1-form. If one starts with a left connection on $\Omega^1$ as is more commonly assumed then applying $\sigma^{-1}$ converts it to a right connection $\nabla$ as above, while dualising it gives a right connection on the space of vector fields which converts to $\hat\nabla$ by application of its $\sigma^{-1}$. We also ask that $\int$ is nondegenerate in the sense $\int ab=0$ for all $a$ requires $b=0$.   Details are in \cite{BegMa:ric}.

\subsection{Case of a parallelisable $\Omega^1$ with basis $\{e^\alpha\}$ and $X(e^\alpha)=X^\alpha$}\label{secpara}

All the models in this paper have $\Omega^1$ parallelisable (a free module) with basis $\{e^\alpha\}$, say. Then partial derivatives $\del_\alpha$ and a related operator $\del^\alpha{}_\beta $ on $A$ are defined by 
\[ \extd f=(\del_\alpha f)e^\alpha,\quad [e^\alpha, f]=\lambda (\del^\alpha{}_\beta f)e^\beta\]
for some parameter $\lambda$ and the former obey a generalised Leibniz rule
\[ \del_\alpha(fg)=(\del_\alpha f)g+ f\del_\alpha g+\lambda(\del_\beta f)\del^\beta{}_\alpha g\]
(here $C_{\alpha\beta}=\delta_{\alpha\beta}+\lambda\del^\alpha{}_\beta$ in the notation of \cite[Sec~1.1]{BegMa}).  The simplest case is when $\{e^\alpha\}$ are central for then $\del^\alpha{}_\beta=0$ and $\del_\alpha$ are derivations on $A$. 

As discussed,  we take classical geodesic-time variable $t$ with $[\extd t,f(t)]=0$, $\extd f(t)=\dot f\extd t$  and a trivial right $B$-bimodule connection $\nabla_B\extd t=0$. Our data for $\nabla_E$ are  a time-dependent function and a left-vector field, which we take as
\[ X^\alpha_t=X_t(e^\alpha)\in A, \quad \kappa_t\in A\]
so that $X_t(\extd f)=X_t((\del_\alpha f)e^\alpha)=(\del_\alpha f)X_t^\alpha$. 

We also fix a right bimodule connection $\nabla$ on $\Omega^1$ expressed as Christoffel symbols defined by
\[ \nabla e^\alpha=-\Gamma^\alpha{}_{\gamma\beta}e^\beta\tens e^\gamma,\]
and typically taken to be a QLC. {\em If $\sigma={\rm flip}$ on the basis} then the auxiliary equations (\ref{auxeqn}) become
\begin{equation}\label{veleq1} [X_t^\alpha,X_t^\beta]=\lambda \del^{[\alpha}{}_\gamma X_t^{\beta]} X_t^\gamma\end{equation}
where $[\alpha...\beta]$ means $\alpha...\beta$ minus the same with $\beta...\alpha$ (otherwise they depend on whatever is the form of $\sigma$). Meanwhile, the velocity equations (\ref{veleq}) applied to a basis element $e^\alpha$ become
\begin{equation}\label{veleq2} \dot X_t^\alpha+ [X_t^\alpha,\kappa_t]-\lambda(\del^\alpha{}_\beta \kappa_t)X_t^\beta+ (\del_\beta X_t^\alpha)X_t^\beta+\Gamma^\alpha{}_{\gamma\beta}(X^\gamma X^\beta+\lambda(\del^\beta{}_\delta X^\gamma) X^\delta) =0.\end{equation}

We also need to assume (in some formal sense) some kind of `integration map'  $\int:A\to \C$ and the conditions (\ref{unitarity}) become under the assumption that $e^\alpha{}^*=e^\alpha$, 
\begin{equation}\label{kappaalpha} \int \kappa_t^* a+ a\kappa_t+(\del_\alpha a)X_t^\alpha=0,\quad \int a X_t^\alpha-X_t^\alpha{}^* a+\lambda (\del^\alpha{}_\beta a)X_t^\beta=0\end{equation}
for all $a$. The divergence condition (\ref{divcond}) becomes 
\[ \int a {\rm div}_{\nabla}(X) + (\del_\alpha a)X^\alpha=0\]
but only needs to be tested on a dual basis $\{f^\alpha\}$. For example, if  $\sigma$ is the flip on the basis and the $\Gamma$ coefficients commute with the basis, we have
\begin{equation}\label{divnablaf} {\rm div}_\nabla(f_\alpha)=\Gamma^\beta{}_{\beta\alpha},\end{equation}
and hence we need
\begin{equation}\label{divcondGamma}  \int a\Gamma^\beta{}_{\beta\alpha}+\del_\alpha a=0\end{equation}
for all $a\in A$ and all $\alpha$. Or if $\nabla=0$ on the basis then it is also zero on the dual basis and we only have the second term. This covers the two cases needed in the paper. Once we have chosen $\int$, the content of (\ref{kappaalpha}) can be  used to solve for $\kappa_t$ and to determine a reasonable reality properties for this and $X^\alpha$. Only after this can we solve (\ref{veleq1})-(\ref{veleq2}) for the vector fields $X_t$. The latest version of \cite{BegMa:ric} gives a slightly more algorithmic approach but requires more machinery whereas the above will be sufficient for our purposes.

Finally, we take $\psi\in E=A\tens B$ (or $\psi_t\in A$ at geodesic time $t$) as our wave function. We then define $\nabla_E$ on $\psi$ by (\ref{nablaE}) and then $\nabla_E\psi=0$ for the `quantum geodesic flow' becomes
\begin{equation}\label{psievol} \dot \psi_t +( \del_\alpha\psi_t) X^\alpha_t + \psi_t \kappa_t=0\end{equation}
which we will solve for a chosen velocity field $X_t$. 

\subsection{At most linear geodesics flows on classical flat  space}\label{secflat}

Geodesics on flat space or spacetime are just straight lines so what could we possibly have to say in this section? The answer is that even in this case we are doing something slightly new, namely a Schroedinger-like `amplitude flow' where $|\psi|^2$ flows like a fluid density. We need to  understand this flat case first before we can consider quantum corrections. The differential calculus is classical with 1-form basis $\extd x^\alpha$ for  coodinates $x^\alpha$. Then the geodesic velocity  equations are just
\begin{equation}\label{veleqnflat} \dot X_t^\alpha +X^\beta_t\del_\beta X_t^\alpha=0.\end{equation}
We can solve this easily in the not totally trivial case where the initial vector field $X$ is at most linear in the $x^\alpha$, so
\begin{equation}\label{Xcu} X_t^\alpha= u^\alpha(t)+ c^\alpha{}_\beta(t) x^\beta\end{equation}
where the coefficients still depend on the geodesic time parameter $t$ (in the Minkowski case, we will denote it $s$). Then (\ref{veleqnflat}) becomes 
\begin{equation}\label{cueqn} \dot c^\alpha{}_\beta+ c^\alpha{}_\gamma c^\gamma{}_\beta=0,\quad  \dot u^\alpha+c^\alpha{}_\beta u^\beta =0.\end{equation}
We also have, assuming everything is real,
\[\nabla\cdot X_t=\Tr(C), \quad \kappa_t=\Tr(C)/2\]
for the unitarity conditions with respect to usual integration.  Finally, for a given $X_t$, we have to solve the `amplitude' geodesic flow equation $\nabla_E\psi=0$, which is just 
\begin{equation}\label{psieqnflat} \dot\psi_t= -X_t^\alpha\del_\alpha\psi_t-  \kappa_t\psi_t.\end{equation}

\begin{proposition}\label{psisol}
At least in the case where the geodesic velocity field $X_t$ is constant or linear in $x^\alpha$, the amplitude flow is 
\[ \psi_t=e^{-\int^t_0\kappa_s \extd s}\psi_0(\phi_t),\quad \phi_t^\alpha(x)= x^\alpha - t X_t^\alpha(x),\quad \phi_t^{-1}{}^\alpha(x)=x^\alpha+ t X_0^\alpha(x).\]
Here $\dot\phi_t^{-1}=X_t(\phi_t^{-1})$ is the velocity along the flow and is independent of $t$, hence equal to the initial value $X_0$. 
\end{proposition}
\proof The solution $\psi_t$ in terms of $\phi_t$ works for any geodesic velocity field $X_t$ and does not need the assumed special form. For the left side of (\ref{psieqnflat}), we have 
\[\frac{\extd \psi_t}{\extd t}=e^{-\int^t_0\kappa_s\extd s}\frac{\partial\psi_0}{\partial \phi_t^\alpha}\frac{\extd \phi_t^\alpha}{\extd t}-\kappa_t\psi_t=e^{-\int^t_0\kappa_s\extd s}\frac{\partial\psi_0}{\partial \phi_t^\alpha}(-X_t^\alpha-t\dot X_t^\alpha)-\kappa_t\psi_t=e^{-\int^t_0\kappa_s\extd s}\frac{\partial\psi_0}{\partial \phi_t^\alpha}(-X_t^\alpha+t X_t^\beta\del_\beta X_t^\alpha)-\kappa_t\psi_t\]
using (\ref{veleqnflat}). For the right side, we have
\begin{align*}-X_t^\alpha\frac{\partial\psi_t}{\partial x^\alpha}-\kappa_t \psi_t&=-X_t^\alpha e^{-\int^t_0\kappa_s\extd s}\frac{\partial\psi_0}{\partial \phi_t^\beta}\frac{\partial\phi_t^\beta}{\partial x^\alpha}-\kappa_t \psi_t=-X_t^\alpha e^{-\int^t_0\kappa_s\extd s}\frac{\partial\psi_0}{\partial \phi_t^\beta}(\delta_{\alpha\beta}-t\del_\alpha X_t^\beta)-\kappa_t \psi_t\end{align*}
using the stated form of $\phi_t^\alpha$, from which we see that (\ref{psieqnflat}) holds. 

For $\phi^{-1}_t(x)$ we proceed in the case of $X_t$ constant plus linear and claim that 
\[ \phi^{-1}_t{}^\alpha(x)=(\id-t C)^{-1}{}^\alpha{}_\beta(x^\beta+t u^\beta).\]
Thus,  
 \begin{align*}\phi^{-1}_t{}^\alpha(\phi_t^\alpha(x))&=(\id-t C)^{-1}{}^\alpha{}_\beta((x^\beta - t X_t^\beta(x))+t u^\beta)=(\id-t C)^{-1}{}^\alpha{}_\beta(x^\beta-c^\beta{}_\gamma x^\gamma t)\\
 &=(\id-t C)^{-1}{}^\alpha{}_\beta (\id-t C)^{-1}{}^\beta{}_\gamma x^\gamma=x\end{align*}
and similarly for $\phi_t(\phi_t^{-1}(x))=x$. Then,  using (\ref{cueqn}), we have 
\begin{align*}
\dot\phi^{-1}_t (x)&=-(\id-t C)^{-1}(-C+tCC)(\id-t C)^{-1}(x+t u)+(\id-t C)^{-1}(u-tCu)\\
&=(\id-t C)^{-1}C(\id-t C)(\id-t C)^{-1}(x+t u)+(\id-t C)^{-1}(\id-t C)u\\
&=(\id-t C)^{-1}C(x+t u)+u\\
&=(\id-t C)^{-1}C(x+t u)+(\id-t C)^{-1}(\id-t C)u\\
&=(\id-t C)^{-1}(Cx+tCu+u-tCu)\\
&=(\id-t C)^{-1}(Cx+u)\\
\ddot\phi^{-1}_t (x)&=(\id-t C)^{-1}CC(x+t u)+(\id-t C)^{-1}(\dot C(x+tu)+C(u+t\dot u))+\dot u\\
&=(\id-t C)^{-1}CC(x+t u)+(\id-t C)^{-1}(-CC(x+tu)+C(u-tCu))-Cu\\
&=(\id-t C)^{-1}C(u-tCu)-Cu\\
&=(\id-t C)^{-1}(\id-tC)Cu-Cu=0
\end{align*}
which means $\dot\phi^{-1}_t$ is independent of  $t$ hence equal to the value at $t=0$. Then $\dot\phi^{-1}_t=C(0)x+u(0)=X_0$ and $\phi_0^{-1}{}^\alpha(x)=x^\alpha$ so that $\phi_t^{-1}{}^\alpha(x)=x^\alpha+ t X_0^\alpha(x)$. Finally,  
\begin{align*}
X_t(\phi_t^{-1})&=X_t((\id-t C)^{-1}{}^\alpha{}_\beta(x^\beta+t u^\beta))\\
&=u+C(\id-tC)^{-1}(x+tu)=(\id-tC)^{-1}(\id-tC)u+(\id-tC)^{-1}(Cx+tCu)\\
&=(\id-tC)^{-1}(u-tCu+Cx+tCu)=(\id-tC)^{-1}(u+Cx)\\
&=\dot\phi^{-1}_t (x)
\end{align*}
as stated. 
 \endproof
 
 We have proven the last part in the constant plus linear case but the same must also hold for general $X_t$ since we know the motion consists of straight lines. The proof in general is not straightforward and would use (\ref{veleqnflat}) in place of the easier (\ref{cueqn}). The same applies to the case of  $X_t=X$ is  time-independent which in the constant plus linear case requires that $C$ must be nilpotent and annihilate $u$, $C^2=0$ and $C.u=0$ so that
 \[ \phi^{-1}_t(x)=x^\alpha+ t u^\alpha+ t C^\alpha{}_\beta x^\beta=x^\alpha+ t X^\alpha=\phi_{-t}(x)\]
as expected since we have a 1-parameter group as per the general theory for the $X$ time-independent case. 

An example of a nontrivial time-dependent family of solutions for (\ref{cueqn}) are
  \begin{equation}\label{excu}
   c_{11}=\frac{1}{t-c},\ c_{12}=0,\ c_{21}=\frac{a}{(t-c)(t-d)},\ c_{22}=\frac{1}{t-d},\  u^1=\frac{b}{t-c},\ u^2=\frac{ab}{(t-c)(t-d)}+\frac{e}{t-d}
  \end{equation}
 where $a,b,c,d,e$ are constant. Here $X_t$ is clearly time-dependent but 
 \[\phi_t^{-1}(x,y)=(x,y)+t X_0(x,y); \quad  X_0(x,y)=(-\frac{b + x}{c},\frac{a  (b+x)-c (y +e )}{c d}), \]
on inverting $\phi_t$ as maps from $\R^2\to \R^2$, has a time-independent effective velocity field $\dot\phi_0^{-1}=X_0$. 

 The interpretation is quite straightforward if the $\R^n$ above is space (rather than spacetime). Then $\psi_t$ is like a wave function over space with time $t$ external to the space. The evolution of $\psi_t$ is a bit like Schroedinger's equation but  everything is compatible with taking $\psi$ real valued. The behaviour of $\rho=|\psi|^2$ should be thought of as a probability or fluid density evolving in time in the sense that the flow equation is like that of a half-density. 

We focussed on the motion of bump functions expressed in $\phi_t^{-1}$ but the actual wave functions in Proposition~\ref{psisol} have a further factor coming from $\kappa_t$. This is a real scaling factor which compensates for the changing density due to the particle motion. For our example (\ref{excu}) it is
\begin{equation}\label{kappafactor}e^{-\int_0^t \kappa_s\extd s}=e^{-{1\over 2}\int_0^t({1\over s-c}+{1\over s-d})\extd s}=\sqrt{cd\over (t-c)(t-d)}\end{equation}
for suitable $c,d$. 
 
\subsection{Minkowski spacetime case} \label{secmink}

In Minkowski space, the geodesics are on spacetime $\R^{n-1,1}$ with time $t$ the last index, so the above geodesic time parameter will now denoted $s$ to avoid confusion. We recall  looking at (\ref{psisol}) that if $\psi_0$ has a bump distribution located at some  $(x,t)$ then the whole distribution including the location of the bump evolves so that the bump at geodesic time $s$ is located at  $\phi_s^{-1}(x,t)$. So for the bump to move, causally we actually want initial points $(x,t)$ such that  $\dot\phi_s^{-1}(x,t)$ are timelike for all $s$, where now dot means with respect to $s$. We saw at least for our constant plus linear class of flows that this is in fact independent of $s$ so the curve is actually a line and  remains timelike.

\begin{figure}
  \begin{minipage}[c]{0.2\textwidth}
    \includegraphics{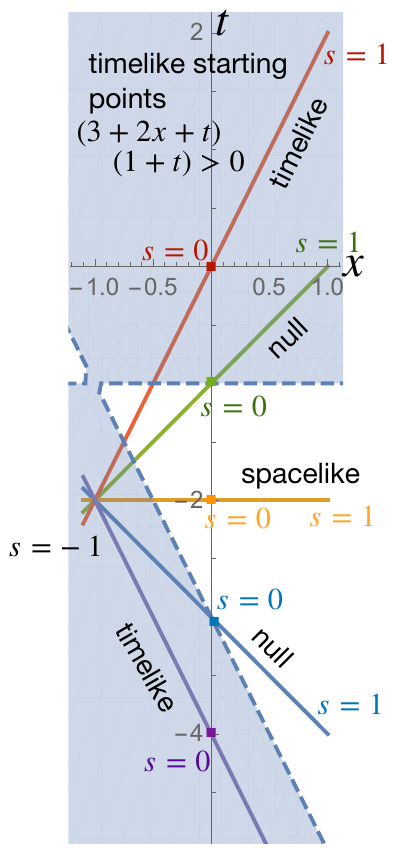}
  \end{minipage}\hfill
  \begin{minipage}[c]{0.7\textwidth}
    \caption{Classical geodesic flows $\phi^{-1}_s(x,t)$ generated by the geodesic vector fields for solution (\ref{excu}) for $a=b=e=1, c=d=-1$ at some initial points $x=0$ and varying $t$ at $s=0$ (shown as nodes). These should be in the shaded region for timelike motion. The world lines pass through the common point shown at $s=-1$. 
    }  \label{figphit}
  \end{minipage}
\end{figure}

We illustrate this in Figure`\ref{figphit}  for an example of our  solutions (\ref{psieqnflat}) but with the geodesic time parameter now changed to $s$. The motion curves $\phi_s^{-1}(x,t)$ are linear as expected but $X_s(x,t)$ is certainly $s$-dependent.  But moving along the flow, $X_s(\phi^{-1}_s(x,t))$ is $s$-independent and in fact coincides with $\dot\phi_s^{-1}(x,t)$. The
condition for the flow to be timelike is for $X_0$ to be, which for our solution (\ref{psieqnflat}) is 
\[ (c (e+t)-(a+d) (b+x)) (c (e+t)-(a-d)(b+x))>0\]
and gives the shaded region in the figure. The different starting points as $s=0$ happen to result in lines that converge at $s=-1$, as shown, which we could read as several world lines at the point $(-1,-2)$ at $s=-1$ evolving from there.  The factor (\ref{kappafactor}) which multiplies the actual wave function solutions is  $1\over {s+1}$ and we see that this blows up at $s=-1$ to reflect this convergence of the geodesic flow from the different starting points. Also note that some of the converging world-lines are  future-oriented, one shown is spacelike and some are past-oriented. An actual physical particle would have a localised form of $\psi_0$ with support in, say, the upper shaded region, so as to be composed of future-pointing ones.

\section{Quantum geodesics on bicrossproduct model flat quantum spacetime}\label{secbic}

We take the $\lambda$-Minkowski quantum spacetime in \cite{MaRue}, where $[x^i,t]=\lambda x^i$ for $i=1,\cdots,n-1$ are the space coordinates and $t$ is the time,  with its $n+1$-dimensional quantum-Poincar\'e invariant calculus\cite{Sit}
\[ [\extd x^i,x^j]=\lambda\delta_{ij}\theta',\quad [\theta',t]=\lambda\theta',\quad [x^i,\extd t]=\lambda \extd x^i,\quad [t,\extd t]=\lambda\theta;\quad \theta=\theta'+\extd t\]
and $[\extd x^i,t]=[\theta',x^i]=0$. Here $[ ,\theta]=\lambda\extd$. This is a $*$-calculus with $x^i,t, \extd x^i,\extd t,\theta'$ all self adjoint with respect to $*$. The quantum metric and flat QLC are\cite[Sec~9.2.2]{BegMa}
\[ g=\extd x^i\tens\extd x^i-\extd t\tens\extd t+\theta\tens\theta,\quad \nabla\extd x^i=\nabla\extd t=\nabla\theta'=0\]
This is presented as a left connection but we can equally view it as a right one since $\sigma={\rm flip}$ on the basis. We do the $n=2$ case and we write $x=x^1, t=x^2$ for brevity, with basis $e^1=\extd x, e^2=\extd t, e^3=\theta'$ of $\Omega^1$, but everything works the same way for higher $n$. 

 \begin{proposition}\label{commutator}
 In 2D flat quantum spacetime where $[x,t]=\lambda x$, we have following commutation relations and exterior derivative.
\begin{align*} 
[x,f(x,t)]&=\lambda x \del_{-\lambda} f(x,t),\quad  [f(x,t),t]=\lambda x {\del \over \del x}f(x,t),\\
[\extd x,f(x,t)]&=\lambda {\del\over\del x}f(x,t+\lambda)\theta',\\
 [f(x,t),\extd t]&=\lambda\left({\del\over\del x}f(x,t)\extd x+\del_{-\lambda} f(x,t)\extd t\right)+\frac{\lambda}{2}\left(\lambda{\del^2\over\del x^2}f(x,t)+(\del_{\lambda}+\del_{-\lambda})f(x,t)+\lambda^2\frac{\del^2}{\del x^2}\del_\lambda f(x,t)\right)\theta'\\
 [\theta',f(x,t)]&=\lambda \del_\lambda f(x,t)\theta',\\
 \extd f(x,t)&={\del\over\del x}f\extd x+\del_{-\lambda} f\extd t+{\lambda\over 2}\Delta_\lambda f\theta'\end{align*}
where $f(x,t)\in A$ is normal ordered so that $t$ is kept to the right and
\[ \del_\lambda f(x,t)={f(x,t+\lambda)-f(x,t)\over\lambda},\quad \Delta_\lambda ={\del^2\over\del x^2}+{1\over\lambda}\left(\del_{-\lambda}-\del_\lambda \right)+\lambda\frac{\del^2}{\del x^2}\del_\lambda.\]
\end{proposition}
\proof 
Setting $f(x,t)=\Sigma_n f_n(x) t^n$, the first formula requires $f_n(x)[x, t^n]=\lambda f_n(x)x\del_{-\lambda}t^n$ for all $f_n$ or 
\[[x, t^n]=x(t^n-(t-\lambda)^n).\]
This follows by induction, since it holds as $[x,t]=x(t-(t-\lambda))=\lambda x$ for $n=1$ and assuming it for $n$, we have 
\begin{align*}[x, t^{n+1}]&=[x, t^n]t+t^n[x,t]=x(t^n-(t-\lambda)^n)t+t^n\lambda x\\
&=x(t^{n+1}-(t-\lambda)^n t)+\lambda(xt^n-x(t^n-(t-\lambda)^n))=x(t^{n+1}-(t-\lambda)^{n+1}).\end{align*}
as required.  We prove the  second, third and the fifth commutation relations using same procedure. For the fourth relation we used induction to prove that
\[ [\extd t, x^n]=-\lambda n x^{n-1}\extd x-\lambda^2 {n(n-1)\over 2}x^{n-2}\theta',\quad  [\extd t, t^n]=((t-\lambda)^n-t^n)\extd t+\frac{1}{2}\left((t-\lambda)^n-(t+\lambda)^n\right)\theta'\]
so that 
\[ [\extd t,f(t)]=-\lambda \del_{-\lambda} f(t)\extd t- \frac{\lambda}{2} (\del_{\lambda}+\del_{-\lambda})f(t)\theta'\]
Setting $f(x,t)=\Sigma_n x^n f_n(t) $, then we get
\begin{align*}
[&\extd t,f(x,t)]=\Sigma_n[\extd t,x^n f_n(t)]=\Sigma_n(x^n[\extd t, f_n(t)]+[\extd t,x^n ]f_n(t))\\
&=-\lambda \del_{-\lambda} f(x,t)\extd t- \frac{\lambda}{2} (\del_{\lambda}+\del_{-\lambda})f(x,t)\theta'+\Sigma_n (-\lambda n x^{n-1}\extd x-\lambda^2 {n(n-1)\over 2}x^{n-2}\theta')f_n(t)\\
&=-\lambda \del_{-\lambda} f(x,t)\extd t- \frac{\lambda}{2} (\del_{\lambda}+\del_{-\lambda})f(x,t)\theta'-\Sigma_n \lambda n x^{n-1}f_n(t)\extd x\\
&\quad -\Sigma_n\lambda^2 {n(n-1)\over 2}x^{n-2}f_n(t)\theta'-\Sigma_n\lambda^2 {n(n-1)\over 2}x^{n-2}\lambda \del_\lambda f_n(t)\theta'\\
&=-\lambda \del_{-\lambda} f(x,t)\extd t- \frac{\lambda}{2} (\del_{\lambda}+\del_{-\lambda})f(x,t)\theta'-\lambda{\del\over\del x}f(x,t)\extd x-\frac{\lambda^2}{2}{\del^2\over\del x^2}f(x,t)\theta'-\frac{\lambda^3}{2}{\del^2\over\del x^2}\del_\lambda f(x,t)\theta'
\end{align*}
which is  answer stated. The exterior derivative is computed from $\extd f(x,t)=-{1\over\lambda}[\theta'+\extd t,f(x,t)]$ and is exhibited as a deformation of the classical $\extd f(x,t)$ plus a correction by a 2nd order operator $\Delta_\lambda$ that deforms  ${\del^2\over\del x^2}+{\del^2\over\del t^2}$. This is similar to calculations in \cite[Chap 9]{BegMa}. \endproof

We will also need the following lemma later.

\begin{lemma}\label{fg} In the above differential calculus on quantum spacetime, let $D f(x)=x{\extd \over\extd x} f(x)$ and $\dot g={\extd \over\extd t}g(t)$. Then
\[ [f(x),g(t)]=\lambda (D f) \dot g- \frac{\lambda^2}{2} (D^2 f)\ddot g+O(\lambda^3)\]
for general functions $f(x),g(t)$. 
\end{lemma}
\proof The derivation will be for polynomials but we also assume the same for other functions in an operator completion of our algebra. We use the above to compute
\begin{align*} [f(x),t^n]&=[f(x),t]t^{n-1}+t[f(x),t]t^{n-2}+\cdots t^{n-1}[f(x),t]\\
&=\lambda (D f)n t^{n-1}-\lambda\left([Df,t]t^{n-2}+[Df,t^2]t^{n-3}+\cdots+[Df,t^{n-1}]\right)\\
&=\lambda (D f){\extd \over\extd t} t^{n}-\lambda^2(D^2 f)(1+2 +\cdots+n-1)t^{n-2}+O(\lambda^3)\\
&=\lambda (D f){\extd \over\extd t} t^{n}-\lambda^2(D^2 f){n(n-1)\over 2}t^{n-2}+O(\lambda^3)\\
&=\lambda (D f){\extd \over\extd t} t^{n}-{\lambda^2\over 2}(D^2 f){\extd^2 \over\extd t^2}t^n+O(\lambda^3)
\end{align*}
where we already know $[Df,t^m]=\lambda D^2 f m t^{m-1}+O(\lambda^2)$ and use this here. This then gives the stated result on general function, including power series where these converge. \endproof

We now proceed to study quantum geodesics. For brevity, we let 
\[ \X(s)=X_s(\extd x)=X_s^1,\quad \Y(s)=X_s(\extd t)=X_s^2,\quad \Theta(s)=X_s(\theta')=X_s^3\]
as elements of $A$, where the geodesic time parameter is now being denoted $s$.  Then the auxiliary equations (\ref{veleq1}) are explicitly
\begin{align}\label{bicvel1} [\X,\Y]&=X_s([\extd x,\Y]-[\extd t,\X]),\quad [\X,\Theta]=X_s([\extd x,\Theta]-[\theta',\X]),\nonumber\\
 [\Y,\Theta]&=X_s([\extd t,\Theta]-[\theta',\Y])\end{align}
which we can further compute using Proposition~\ref{commutator}. For the velocity equation (\ref{veleq2}), we then similarly have
\begin{align}\label{bicvel2} \dot \X&+[\X,\kappa_s]+X_s(\extd \X-[\extd x,\kappa_s])=0,\quad \dot \Y+ [\Y,\kappa_s]+X_s(\extd \Y-[\extd t,\kappa_s])=0,\nonumber\\
&  \dot \Theta+[\Theta,\kappa_s]+X_s(\extd \Theta-[\theta',\kappa_s])=0\end{align}
which again can be computed further using the commutation relations.

\begin{lemma} We define $\int f(x,t)=\int\extd x\extd t\, f(x,t)$ for $f$ on the left a normal ordered function of $x,t\in A$ and $f$ on the right the same function integrated classically. Then the condition (\ref{divcondGamma}) for the two divergences to agree  formally holds on a reasonable class of functions $f$. 
\end{lemma}
\proof Here $\Gamma^\alpha{}_{\beta\gamma}=0$ we we just need that $\int \del_\alpha f(x,t)=0$. This is clear for $\alpha=1$ as $\del_1={\del\over\del x}$ on normal ordered functions. For $\del_2=\del_\lambda$ on normal ordered functions and the condition amounts to $\int\extd t\, f(t+\lambda)=\int\extd t\, f(t)$ for any function $f(t)$. In fact $\lambda$ is imaginary so this is not merely a change of variables but we assume that our class of functions are such that this makes sense and formally holds. Then $\int\Delta_\lambda f=0$ as a consequence of the first two cases as well as $\int\extd t {\del \over\del t} f(t)=0$. \endproof

It remains to exhibit some solutions for $\X,\Y,\Theta\in A$ (as normal ordered functions of $x,t$) for some reasonable $\kappa_s$ and for suitable reality assumptions so as to obey (\ref{kappaalpha}). After this, we have to solve  the $\nabla_E\psi=0$ equation for $\psi_s\in A$ at each $s$, namely
\begin{equation}\label{bicflow} {\del \psi_s\over\del s}=-X_s(\extd \psi_s)-\psi_s\kappa_s.\end{equation}
We will do this for the simplest two cases, namely $X,Y,\Theta$ constant or constant plus linear on spacetime.

\subsection{Solutions with $\X,\Y,\Theta=0$ constant on spacetime} \label{secon}

The most trivial case, which corresponds as we shall see, to linear motion, is to suppose that $\X,\Y,\Theta,\kappa\in \C.1$ i.e. multiples of the identity or constants in $A$ (but still potentially geodesic-time dependent). Then the auxiliary equation (\ref{veleq1}) becomes empty and the velocity equation (\ref{veleq2}) is
\[ \dot \X=0,\quad \dot \Y=0,\quad \dot\Theta=0\]
so these are also constants in $s$. Moreover, the first half of the unitarity condition (\ref{kappaalpha}) says that $\kappa$ is imaginary and we take  
\[ \kappa=0\]
since the effect of a purely imaginary constant is just to add a phase in the evolution of $\psi$.  The second half of (\ref{kappaalpha}) says that $\X,\Y,\Theta$ are real given the following result.

\begin{lemma} For the above model, $\int(\del^\alpha{}_\beta a) X^\beta=0$ holds on a reasonable class of functions $a$.
\end{lemma}
\proof We have $\int(\lambda\del^\alpha{}_\beta a) X^\beta=\int X_t[e^\alpha,a]$. For $\alpha=1$,we have\[ \int X_t[\extd x,a]=\int \lambda \frac{\partial}{\partial x}a(x,t+\lambda)\Theta=\lambda\Theta\int \frac{\partial}{\partial x}a(x,t+\lambda)=0\] 
For $\alpha=2$,we have\[\int X_t[\extd t,a]=\int -\lambda\left(\X{\del\over\del x}a(x,t)+\Y\del_{-\lambda} a(x,t)\right)-\frac{\lambda}{2}\left(\lambda{\del^2\over\del x^2}a(x,t)+(\del_{\lambda}+\del_{-\lambda})a(x,t)+\lambda\frac{\del^2}{\del x^2}\del_\lambda a(x,t)\right)\Theta=0\]
For $\alpha=3$,we have\[ \int X_t[\theta',a]=\int \lambda \partial_\lambda a(x,t)\Theta=\Theta \lambda\int \partial_\lambda a(x,t)=0\] 
using the translation invariance in the previous lemma. 
 \endproof

Assuming so, the equation for normal ordered and geodesic time-dependent elements $\psi_s(x,t)$ is
\begin{align}\label{solution}
{\del \psi\over\del s}=-({\del\over\del x}\psi) \X- (\del_{-\lambda} \psi)\Y-{\lambda\over 2}(\Delta_\lambda\psi )\Theta
\end{align}
which in one extreme
\[ \X=\Y=0,\quad  \Theta=- {\hbar\over m  \lambda_P}\]
is a deformed Schroedinger equation on $\R^2$ where one of the directions has a finite-difference derivative. Another extreme, which we focus on, is to take $\Theta=0$ then 
\begin{equation}\label{soltheta0}\psi_s(x,t)=e^{-s\X{ \del\over\del_x}-s\Y\del_{-\lambda}} \psi_0(x,t)=e^{-s\Y\del_{-\lambda}} \psi_0(x-s\X,t)\end{equation}
on normal ordered functions. In the classical case where $\del_{-\lambda}={\del\over\del t}$, this would be 
\begin{equation}\label{psic} \psi^c_s(x,t)=\psi_0^c(x-s\X,t-s\Y)\end{equation}
or linear motion on the underlying spacetime, with world line slope determined by the constants $\X,\Y$. We could also bring in $\Theta\ne 0$ to give Schroedinger equation-like corrections to this according to $\Delta_\lambda$. 

\subsubsection{Example of plane wave initial amplitude}  \label{secconwave}

We keep to $\Theta=0$ for simplicity and compute the flow (\ref{soltheta0}) for an initial plane wave $\psi_0(x,t)=e^{- \imath k x}e^{\imath \omega t}$ at $s=0$. We have
\[ \del_{-\lambda} e^{\imath\omega t}={e^{\imath\omega (t-\lambda)}-e^{\imath\omega t}\over-\lambda}=\left({e^{-\imath {\lambda}\omega}-1\over -\lambda}\right)e^{\imath\omega t}=\imath\omega\left({e^{\lambda_p\omega }-1\over \lambda_p\omega}\right)e^{\imath\omega t}\]
from which it immediately follows that 
\begin{align*}
 \psi_s(x,t)&=e^{- \imath k (x- s \X)} e^{-s \Y\imath\omega \left({e^{\lambda_p\omega }-1\over \lambda_p\omega}\right)}e^{\imath\omega t}= e^{- \imath k (x- s \X)}e^{\imath \omega\left( t-s \Y({e^{\lambda_p\omega}-1\over \lambda_p\omega})\right)}\\
 &=\psi_0\left(x-s\X, t-s \Y({e^{\lambda_p\omega}-1\over \lambda_p\omega})\right)  \end{align*}
which we see keeps the form of the initial normal ordered plane wave but for a displaced origin. Moreover, compared to the classical flow (\ref{psic}), the displaced orging moves with $\Y$ changed  to 
\[ \Y'=\Y({e^{\lambda_p\omega}-1\over \lambda_p\omega})\]
which in turn means the classical geodesic velocity $v=\X/\Y$ is changed to 
\[{\X\over\Y'}=v {\lambda_p\omega \over  e^{\lambda_p\omega }-1}=v\left(1-\frac{\lambda_p\omega}{2}+\frac{(\lambda_p\omega)^2}{12}-\frac{(\lambda_p\omega)^4}{720}+\cdots\right).\]
 A classical geodesic just shifts the origin of the coordinate system in a line as $s$ increases, with velocity $v$, and the effect of the quantum spacetime as far as the form of the normal ordered plane wave is concerned is to amend that as shown by an amount that depends on the wave frequency. This is broadly in agreement with heuristic arguments about a `variable speed of light' in \cite{AmeMa} but now with a precise operational meaning. This is for one plane-wave mode but by linearity the same applies for every frequency $\omega$ of a wave packet. 

We are also free to explore expectation values according to our interpretation of $\psi$ as a wave-function over spacetime (and probabilities understood with respect to the geodesic time $s$), which is more physical than looking at the form of the normal ordered plane wave. Clearly, with the partial derivatives defined as the usual ones on normal ordered functions, so for a single plane wave
\[ \<\del_x\>={\int \psi_s^*\del_x\psi\over \int \psi^*\psi}=-\imath k,\quad \<\del_t\>={\int \psi_s^*\del_t\psi\over \int \psi^*\psi}=\imath \omega\]
by the same calculation as classically (ignoring as usual that planes waves are not normalisable and cancelling the infinite $\int \psi_s^*\psi_s=\int 1$ in the ratio). More interesting using the commutation relations and the same cancellations is
\[ \<x\>= e^{-\lambda_p\omega} s \X,\quad \<t\>= s \Y \left({e^{\lambda_p\omega}-1\over \lambda_p\omega}\right)- \lambda_p  k \<x\>.\]
Here, in making sense of the infinite ratios, we should change variables to $t'=t-s \Y'$ and $x'=x-s \X$ since it is these that are in the plane wave exponents whereby $\int \extd x' \extd t'=\int 1$ etc. Then,
\begin{align*} \<x\>&=(\int 1)^{-1}\int e^{-\imath\omega t'}xe^{\imath\omega t'}=(\int 1)^{-1}\int (x+[e^{-\imath \omega t'},x]e^{\imath\omega t'})= (\int 1)^{-1}\int (x + (e^{\imath\lambda\omega}-1)x)\\
&= e^{-\lambda_p\omega} \int (x'+s\X)= e^{-\lambda_p\omega}s\X
\end{align*}
where $\int x'=0$ for the real part of plane waves centred at the origin as they are in the $x',t'$ coordinates. Similarly,
\begin{align*}
\<t\>&=(\int 1)^{-1}\int e^{-\imath\omega t'}e^{\imath k x'}t e^{-\imath k x'}e^{\imath\omega t'}=(\int 1)^{-1}\int (t+e^{-\imath\omega t'}e^{\imath k x'}[t,e^{-\imath kx'}]e^{\imath\omega t'})\\
&=(\int 1)^{-1}\int (t+e^{-\imath\omega t'}\imath k \lambda x e^{\imath\omega t'})=(\int 1)^{-1}\int (t+\imath k \lambda x -\lambda \imath k[x,e^{-\imath\omega t'}]e^{\imath\omega t'})\\
&=(\int 1)^{-1}\int (t+\imath k \lambda x+\imath k\lambda(e^{\imath\omega\lambda}-1)x)=(\int 1)^{-1}\int (t'+s\Y'-k \lambda_p e^{-\lambda_p\omega}x)\\
&=s\Y'-  \lambda_p ke^{-\lambda_p\omega}s\X
\end{align*}
where again we assume $\int t'=0$ for the real part of plane waves centred at the origin as they are in $x',t'$ coordinates. The assignment of the integrals here is schematic and should be understood more precisely by working with wave packets, in a routine manner. 

To understand this a little more, we put our initial plane wave on shell for a massless particle, say. 
In fact, we only need the classical value for the lowest order correction and take $k=\omega$. Then we have an effective geodesic velocity from the  point of view of expectation values is
\[{\<x\>\over\<t\>}\approx  v\left(1+\lambda_p\omega(v-{3\over 2})+\cdots\right) \]
where $v=\X/\Y$ as before. This is different from the answer we obtained from the form of the normal plane waves unless $v=1$, but more physical.  For small geodesic velocities our correction is three times the amount but the same sign as that suggested by the normal ordered form. 

\subsection{Perturbative approach to the amplitude flow} 

To solve (\ref{solution}) in $\Theta=0$ more generally, we use perturbation theory. We first consider the equation to order $\lambda$, 
\begin{align}\label{solution2}
{\del \psi\over\del s}=-({\del\over\del x}\psi )\X- ({\del\over\del t}\psi )\Y+\frac{\lambda}{2}({\del^2\over\del t^2}\psi )\Y
\end{align}
We define $\psi^1 $ such that $\psi =\psi^c +\lambda\psi^1 $, where $\psi^c $ are the solution of the classical case.  Then $\psi^1 $ obeys 
\begin{align*}
{\del \psi^1\over\del s}&=-({\del\over\del x}\psi^1 )\X- ({\del\over\del t}\psi^1 )\Y+\frac{1}{2}({\del^2\over\del t^2}\psi^c )\Y\end{align*}
after divding through by $\lambda$. For convenience we assume here that $\X,\Y$ are constant as elements of $A$ and do not depend on $\lambda$ (we return to the more general case later). One can likewise proceed perturbatively to higher order, thus $\psi=\psi^c + \lambda^1\psi^1+\lambda^2\psi^2$ for the next order correction. Putting this into (\ref{solution}) and dividing through by $\lambda^2$, we  find \begin{align*}
{\del \psi^2\over\del s}&=-({\del\over\del x}\psi^2 )\X- ({\del\over\del t}\psi^2 )\Y+\frac{1}{2}({\del^2\over\del t^2}\psi^1 )\Y-\frac{1}{6}({\del^3\over\del t^3}\psi^c )\Y
\end{align*}
Similarly to any order. 

Proceeding with this scheme and $\X,\Y$ constant, we have given $\psi^c$ in (\ref{psic}) and now write $\psi_0^c=\psi_0^0$ to remind us that this is the lowest term in our order $\lambda$ expansion. We also adopt the shorthand
\begin{equation}\label{uv} u=x-s\X,\quad v=t-s\Y.\end{equation}
Then we have 
\begin{align*} \psi^c_s(x,t)&=\psi_0^0(u,v)\\
 \psi^1_s(x,t)&=\psi_0^1(u,v)+\frac{1}{2}s\Y\frac{\del^2 \psi_0^0(u,v)}{\del t^2}\\
\psi^2_s(x,t)&=\psi_0^2(u,v)+\frac{1}{2}s\Y\frac{\del^2 \psi_0^1(u,v)}{\del t^2}-\frac{1}{6}s\Y\frac{\del^3 \psi_0^0(u,v)}{\del t^3}+\frac{1}{8}s^2 \Y^2\frac{\del^4 \psi_0^0(u,v)}{\del t^4}
\end{align*}

\begin{proposition}  For constant $\X,\Y$ and $\Theta=0$, an order by order perturbative solution \[ \psi_s(x,t)=\sum_{n=0}^\infty \lambda^n\psi^n_s(x,t)\] of (\ref{solution})  requires at order $n$ that 
\begin{align}\label{psineqn}
\frac{\del\psi^n}{\del s}=-\frac{\del\psi^n}{\del x}\X+\sum\limits_{m=0}^{n}\frac{(-1)^{n+1-m}}{(n+1-m)!}\frac{\del^{n+1-m}}{\del t^{n+1-m}}\psi^m \Y \end{align}
and has general solution
\[ \psi_s^n(x,t)=\psi_0^n(u,v)+ \sum\limits_{i=1}^n \sum\limits_{j=1}^i c_{ij}(s\Y)^j\frac{\del^{i+j} \psi_0^{n-i}(u,v)}{\del t^{i+j}}\]
where $c_{ij}$ are defined iteratively by
\[c_{i1}=\frac{(-1)^{i+1}}{(i+1)!},\quad c_{i,k}=\frac{1}{k}\sum\limits_{j=1}^{j=i-k+1}\frac{(-1)^{j+1}}{(j+1)!}c_{i-j,k-1}, \quad  k=2,\cdots,i.\]
 \end{proposition}
\proof
In our $\Theta=0$ case, (\ref{solution}) reduces to
\[{\del \psi\over\del s}=-({\del\over\del x}\psi) \X+ \frac{1}{\lambda}\left( \psi(x,t-\lambda)-\psi (x,t)\right)\Y=-({\del\over\del x}\psi) \X-\sum_{m=1}^\infty\frac{(-\lambda)^{m-1}}{m!}\frac{\del^m \psi}{\del t^m}\Y.\] 
Substituting $\psi_s(x,t)=\sum_{n=0}^\infty \lambda^n\psi^n_s(x,t)$ into this gives\begin{align*}
\sum_{n=0}^\infty \lambda^n {\del \psi^n\over\del s}&=-\sum_{n=0}^\infty \lambda^n {\del \psi^n\over\del x}\X-\sum_{n=0}^\infty \sum_{m=1}^\infty\frac{(-1)^{m-1}\lambda^{n+m-1}}{m!}\frac{\del^m \psi^n}{\del t^m}\Y \\
&=-\sum_{n=0}^\infty \lambda^n {\del \psi^n\over\del x}\X-\sum_{n=0}^\infty \sum_{j=n}^\infty\frac{(-1)^{j-n}\lambda^{j}}{(j+1-n)!}\frac{\del^{j+1-n} \psi^n}{\del t^{j+1-n}}\Y\\
&=-\sum_{n=0}^\infty \lambda^n {\del \psi^n\over\del x}\X-\sum_{j=0}^\infty \sum_{n=0}^j\frac{(-1)^{j-n}\lambda^{j}}{(j+1-n)!}\frac{\del^{j+1-n} \psi^n}{\del t^{j+1-n}}\Y\\
&=-\sum_{n=0}^\infty \lambda^n {\del \psi^n\over\del x}\X-\sum_{n=0}^\infty \sum_{m=0}^n\frac{(-1)^{n-m}\lambda^{n}}{(n+1-m)!}\frac{\del^{n+1-m} \psi^m}{\del t^{n+1-m}}\Y
\end{align*}
giving (\ref{psineqn}) when we solve order by order. This then leads to the recursion relations as stated. 
\endproof

It is a useful check to make sure we agree with the non-perturbative plane wave results in the preceding section. Here we set $\psi^0_0=e^{- \imath k x}e^{\imath \omega t}$. Then 
\begin{align*}\psi^1&=-\frac{1}{2}s\omega^2\Y e^{- \imath k (x-s\X)}e^{\imath \omega (t-s\Y)}=-\frac{1}{2}s\omega^2\Y\psi^c\\
 \psi^2&=\frac{1}{24}(4\imath+3s\omega \Y)s\omega^3 \Y e^{- \imath k (x-s\X)}e^{\imath \omega (t-s\Y)}=\frac{1}{24}(4\imath+3s\omega \Y)s\omega^3\Y\psi^c\\
 \psi^3&=-\frac{1}{48}(-2+4\imath s\omega \Y+s^2\omega^2 \Y^2)s\omega^4\Y e^{- \imath k (x-s\X)}e^{\imath \omega (t-s\Y)}=-\frac{1}{48}(-2+4\imath s\omega \Y+s^2\omega^2 \Y^2)s\omega^4\Y\psi^c
 \end{align*}
 etc., so that 
\begin{align*}\psi_s&=\left(1-  \frac{\lambda}{2}s\omega^2\Y+ \frac{\lambda^2}{24}(4\imath+3s\omega \Y)s\omega^3\Y-\frac{\lambda^3}{48}(-2+4\imath s\omega \Y+s^2\omega^2 \Y^2)s\omega^4\Y+\cdots \right)e^{- \imath k (x-s\X)}e^{\imath \omega (t-s\Y)}\\
&= e^{\imath s\omega \Y {-\lambda_p\omega\over 2} \left(1+\frac{\lambda_p\omega}{3}+\frac{(\lambda_p\omega)^2}{12}+\cdots\right)}e^{- \imath k (x- s \X)}e^{\imath \omega( t- s \Y)}\\
&= e^{\imath s\omega \Y \left(1+ {e^{\lambda_p\omega}-1\over -\lambda_p\omega}\right)}e^{- \imath k (x- s \X)}e^{\imath \omega( t- s \Y)}= e^{- \imath k (x- s \X)}e^{\imath \omega\left( t-s \Y({1- e^{\lambda_p\omega}\over -\lambda_p\omega})\right)}
\end{align*}
which we see agrees with the expansion of the exact answer.  We made a polar decomposition of the bracketted expression -- it turns out to be the required phase -- and then identified this with the expansion of the phase shift in the exact answer as shown.

\subsubsection{Example of Gaussian initial amplitudes} We now relate to some physics such as $\lambda$-dependent effective speed of light was predicted in the naive `plane wave ansatz' in the 1990s and approach this more scientifically via quantum geodesics. We consider the case where $\psi^n_0=0$ for $n>0$ (so initially, no quantum correcton). Then the general perturbative solution becomes
 \[ \psi_s^n(x,t)=\sum\limits_{j=1}^n c_{nj}(s\Y)^j\frac{\del^{n+j} \psi_0^0(u,v)}{\del t^{n+j}}\]
 If initially $\psi^0_0$ is a Gaussian centred at $x=t=0$ which means $\psi^0_0=e^{-\frac{x^2}{\beta}}e^{-\frac{t^2}{\beta}}$, then we have 
  \[\psi^1=e^{-\frac{1}{\beta}(x-s\X)^2}e^{-\frac{1}{\beta}(t-s\Y)^2}\frac{s\Y}{\beta^2}[-\beta+2(t-s\Y)^2]=\psi^c \frac{s\Y}{\beta^2}[-\beta+2(t-s\Y)^2]\]
  \begin{align*}
  \psi^2&=e^{-\frac{1}{\beta}(x-s\X)^2}e^{-\frac{1}{\beta}(t-s\Y)^2}\frac{s\Y}{6\beta^4}[12s\Y(t-s\Y)^4+4(2t-11s\Y)(t-s\Y)^2\beta+3(-4t+7s\Y)\beta^2]\\
  &=\psi^c\frac{s\Y}{6\beta^4}[12s\Y(t-s\Y)^4+4(2t-11s\Y)(t-s\Y)^2\beta+3(-4t+7s\Y)\beta^2]
  \end{align*}
  Thus \[\psi_s=\psi^c\left\{1+\lambda \frac{s\Y}{\beta^2}[-\beta+2(t-s\Y)^2]+\lambda^2 \frac{s\Y}{6\beta^4}[12s\Y(t-s\Y)^4+4(2t-11s\Y)(t-s\Y)^2\beta+3(-4t+7s\Y)\beta^2]+\cdots \right\}\]
  where the dots are $O(\lambda^3)$. Working to this order, we now define $f_1, f_2$ such that $\psi^1=\psi^c f_1, \psi^2=\psi^c f_2$ and find
\begin{align*}
\<a\>&=\frac{\int\psi^* a \psi}{\int\psi^* \psi} \approx\frac{\int(1-\lambda f_1+\lambda^2 f_2)\psi^{c*} a \psi^c (1+\lambda f_1+\lambda^2 f_2)}{\int(1-\lambda f_1+\lambda^2 f_2)\psi^{c*} \psi^c(1+\lambda f_1+\lambda^2 f_2)}\\
& \approx\frac{\int\psi^{c*} a \psi^c+\lambda [\psi^{c*} a \psi^c, f_1]+\lambda^2 \psi^{c*} a \psi^c(2f_2-f_1^2)}{\int\psi^{c*} \psi^c+\lambda [\psi^{c*} \psi^c, f_1]+\lambda^2 \psi^{c*}  \psi^c(2f_2-f_1^2)}\\
&\approx\frac{\int\psi^{c*} a \psi^c+\lambda^2 D(\psi^{c*} a \psi^c) \dot f_1+\lambda^2 \psi^{c*} a \psi^c(2f_2-f_1^2)}{\int\psi^{c*} \psi^c+\lambda^2 D(\psi^{c*} \psi^c) \dot f_1+\lambda^2 \psi^{c*}  \psi^c(2f_2-f_1^2)}\\
&\approx\frac{\int\psi^{c*} a \psi^c+\lambda^2 D(\psi^{c*} a \psi^c) \dot f_1+\lambda^2 \psi^{c*} a \psi^c(2f_2-f_1^2)}{\int\psi^{c*} \psi^c},
\end{align*}
where we use Lemma \ref{fg} in the third line and definition of  $f_1, f_2$ in the fourth line. To look more explicitly, we let  $x_s=x-s\X,t_s=t-s\Y$ be a shorthand. Then
\begin{align*}
\<x\>&\approx\frac{\int\psi^{c*} x \psi^c+\lambda^2 D(\psi^{c*} x \psi^c) \dot f_1+\lambda^2 \psi^{c*} x \psi^c(2f_2-f_1^2)}{\int\psi^{c*} \psi^c}\\
&\approx\frac{\int\psi^{c*} x \psi^c+\int\lambda^2 x\frac{\del}{\del x}(x e^{-\frac{2 x_s^2}{\beta}} e^{-\frac{2 t_s^2}{\beta}}) \frac{4s\Y}{\beta^2}(t-s\Y)+\int\lambda^2 x e^{-\frac{2 x_s^2}{\beta}} e^{-\frac{2 t_s^2}{\beta}}(2f_2-f_1^2)}{\int\psi^{c*} \psi^c}\\
&=\frac{\int\psi^{c*} x \psi^c}{\int\psi^{c*} \psi^c}=\frac{\int e^{-\frac{t_s^2}{\beta}}e^{-\frac{x_s^2}{\beta}} x e^{-\frac{x_s^2}{\beta}}e^{-\frac{t_s^2}{\beta}}}{\int e^{-\frac{t_s^2}{\beta}}e^{-\frac{2 x_s^2}{\beta}} e^{-\frac{t_s^2}{\beta}}}=\frac{\int x e^{-\frac{2 x_s^2}{\beta}} e^{-\frac{2 t_s^2}{\beta}}+[e^{-\frac{t_s^2}{\beta}},xe^{-\frac{2 x_s^2}{\beta}}]e^{-\frac{t_s^2}{\beta}}}{\int  e^{-\frac{2 x_s^2}{\beta}} e^{-\frac{2 t_s^2}{\beta}}+[e^{-\frac{t_s^2}{\beta}}, e^{-\frac{2 x_s^2}{\beta}}]e^{-\frac{t_s^2}{\beta}}}\\
&\approx\frac{\frac{1}{2}\pi s\X\beta+\int\left(-\lambda D(x e^{-\frac{2 x_s^2}{\beta}})\frac{\extd}{\extd t}e^{-\frac{t_s^2}{\beta}}+\frac{\lambda^2}{2}D^2(x e^{-\frac{2 x_s^2}{\beta}})\frac{\extd^2}{\extd t^2}e^{-\frac{t_s^2}{\beta}}\right)e^{-\frac{t_s^2}{\beta}}}{\frac{1}{2}\pi\beta+\int\left(-\lambda D( e^{-\frac{2 x_s^2}{\beta}})\frac{\extd}{\extd t}e^{-\frac{t_s^2}{\beta}}+\frac{\lambda^2}{2}D^2( e^{-\frac{2 x_s^2}{\beta}})\frac{\extd^2}{\extd t^2}e^{-\frac{t_s^2}{\beta}}\right)e^{-\frac{t_s^2}{\beta}}}\\
&=\frac{\frac{1}{2}\pi s\X\beta-\frac{1}{4}\pi s\X\lambda^2}{\frac{1}{2}\pi\beta}=s\X\left(1-\frac{\lambda^2}{2\beta}\right)
\end{align*}
where used the value of  $\dot f_1$ for the second line and similarly the value of $2f_2-f_1^2$ for the third line. Because the integrals involving these already have $\lambda^2$ in front, we can treat them classically and find that they vanish. 
\begin{align*}
\<t\>&\approx\frac{\int\psi^{c*} t \psi^c+\lambda^2 D(\psi^{c*} t \psi^c) \dot f_1+\lambda^2 \psi^{c*} t \psi^c(2f_2-f_1^2)}{\int\psi^{c*} \psi^c}\\
&\approx\frac{\int e^{-\frac{t_s^2}{\beta}}e^{-\frac{x_s^2}{\beta}} t e^{-\frac{x_s^2}{\beta}}e^{-\frac{t_s^2}{\beta}}+\frac{-3}{4}\lambda^2 \pi s\Y}{\frac{1}{2}\pi\beta}=\frac{\int e^{-\frac{t_s^2}{\beta}}e^{-\frac{2 x_s^2}{\beta}} t e^{-\frac{t_s^2}{\beta}}+e^{-\frac{t_s^2}{\beta}}e^{-\frac{ x_s^2}{\beta}} [t,e^{-\frac{ x_s^2}{\beta}}] e^{-\frac{t_s^2}{\beta}}+\frac{-3}{4}\lambda^2 \pi s\Y}{\frac{1}{2}\pi\beta}\\
&\approx\frac{\int e^{-\frac{2 x_s^2}{\beta}} e^{-\frac{2 t_s^2}{\beta}} t +[e^{-\frac{ t_s^2}{\beta}},e^{-\frac{ 2 x_s^2}{\beta}}]e^{-\frac{ t_s^2}{\beta}} t-\lambda e^{-\frac{t_s^2}{\beta}}e^{-\frac{ x_s^2}{\beta}}D(e^{-\frac{ x_s^2}{\beta}}) e^{-\frac{t_s^2}{\beta}}+\frac{-3}{4}\lambda^2 \pi s\Y}{\frac{1}{2}\pi\beta}\\
&=\frac{\int e^{-\frac{2 x_s^2}{\beta}} e^{-\frac{2 t_s^2}{\beta}} t +[e^{-\frac{ t_s^2}{\beta}},e^{-\frac{ 2 x_s^2}{\beta}}]e^{-\frac{ t_s^2}{\beta}} t-\lambda e^{-\frac{ x_s^2}{\beta}}D(e^{-\frac{ x_s^2}{\beta}}) e^{-\frac{2 t_s^2}{\beta}}-\lambda[e^{-\frac{t_s^2}{\beta}},e^{-\frac{ x_s^2}{\beta}}D(e^{-\frac{ x_s^2}{\beta}}) ]e^{-\frac{t_s^2}{\beta}}+\frac{-3}{4}\lambda^2 \pi s\Y}{\frac{1}{2}\pi\beta}\\
&\approx\frac{\int e^{-\frac{2 x_s^2}{\beta}} e^{-\frac{2 t_s^2}{\beta}} t +\left(-\lambda D( e^{-\frac{2 x_s^2}{\beta}})\frac{\extd}{\extd t}e^{-\frac{t_s^2}{\beta}}+\frac{\lambda^2}{2}D^2( e^{-\frac{2 x_s^2}{\beta}})\frac{\extd^2}{\extd t^2}e^{-\frac{t_s^2}{\beta}}\right)e^{-\frac{ t_s^2}{\beta}} t-\lambda e^{-\frac{ x_s^2}{\beta}}D(e^{-\frac{ x_s^2}{\beta}}) e^{-\frac{2 t_s^2}{\beta}}+\frac{-3}{4}\lambda^2 \pi s\Y}{\frac{1}{2}\pi\beta}\\
&=\frac{\int e^{-\frac{2 x_s^2}{\beta}} e^{-\frac{2 t_s^2}{\beta}} t-\lambda^2 \pi s\Y}{\frac{1}{2}\pi\beta}=\frac{\frac{1}{2}\pi\beta s\Y-\lambda^2 \pi s\Y}{\frac{1}{2}\pi\beta}=s\Y \left(1-\frac{2\lambda^2}{\beta}\right)
\end{align*}
Thus, the effective velocity is
\begin{align*}
\frac{\<x\>}{\<t\>}=\frac{\X}{\Y}\frac{\left(1-\frac{\lambda^2}{2\beta}\right)}{\left(1-\frac{2\lambda^2}{\beta}\right)}\approx {\X\over\Y}\left(1+\frac{3\lambda_p^2}{2\beta}\right)
\end{align*}
as a factor times the classical velocity. We have a different answer than from plane waves, i.e. the correction to the effective geodesic velocity depends on the shape of the initial wave function being geodesically evolved. 

\begin{figure}
\[ \includegraphics[scale=0.9]{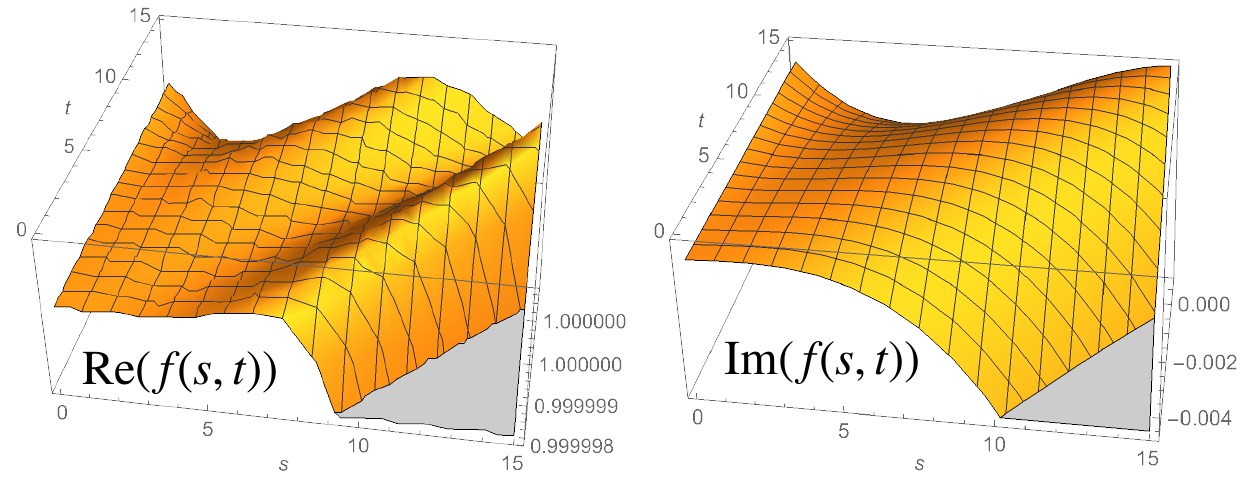}\]
\caption{ Real and imaginary parts of distortion factor  $f(s,t)$ in the evolution of a Gaussian  with  $v_t=1,\beta=20,\lambda_p=0.001$. \label{plotpsi}}
\end{figure}

We also see that the Gaussian changes shape as it evolves. To exhibit this, we let 
\[ \psi_s(x,t)=\psi^c_s(x,t) f(s,t),\quad \psi^c_s(x,t)=e^{-\frac{( x-s\X)^2}{\beta}}e^{-\frac{(t-s\Y)^2}{\beta}}\]
and plot real and imaginary parts of the distortion factor $f(s,t)$  in Figure~\ref{plotpsi}, with $\Y=1,\beta=20,\lambda_p=0.001$. A constant value $f=1$ would mean that the initial Gaussian keeps its
form but the centre merely moves to along a line in spacetime as $s$ evolves. In the figure on the left the central valley at $t=s$ (since $\Y=1$) is the location of the centre of the Gaussian but we see that there is a fluctuation on either side of this, and in particular that we acquire an imaginary part as $s$ increases. 

\subsection{Solutions with $\X,\Y,\Theta$ linear or constant on spacetime}\label{seclin}

Here we extend the above ansatz for $\X,\Y,\Theta$ to include linear dependence on $x,y$ so
\[ \X=u_1 1+ c_{11} x+c_{12}t,\quad \Y=u_2 1+ c_{21} x+ c_{22}t,\quad \Theta= w 1+ c_1 x+ c_2 t\]
say and analyse the above for ($s$-dependent) coefficients $u_i,c_{ij},c_i, w\in \C.1$ (i.e. constant in $x,t$). We regard $C=(c_{ij})$ as a matrix as we did before in Section~\ref{secflat}. Noting that $\del_\lambda t=1$ is the same as classically, we see that $\Delta_\lambda=0$ and hence $\del_3=0$ on the above variables and that
\[ \del_i \X=c_{1i},\quad  \del_i \Y=c_{2i},\quad \del_i\Theta=c_i\]
for $i=1,2$.  

\begin{lemma} \label{bicuni} For (\ref{kappaalpha}) to hold, we need 
\begin{align}\label{newunitarity}
\overline{c_{ij}}=c_{ij},\quad \overline{c_{i}}=c_{i},\quad u_1-\overline{u_1}=\lambda(c_{12}+c_1),\quad u_2-\overline{u_2}=-\lambda (c_{11}+c_2),\quad w-\overline{w}=2\lambda c_2
\end{align}

\end{lemma}
\proof For the second half of (\ref{kappaalpha})  we need $\int aX^\alpha-X^{\alpha *} a+X[e^\alpha,a]=0$. We will approach this formally under the assumption that test functions $a$ decay suitably at infinity (so that we can cancel a total divergence) and are such that $\int a(x,t+\lambda)=\int a(x,t)$ as an extension of translation invariance on normal ordered functions to imaginary displacements. In this case, we have
\begin{align*}\int ax &=\int xa-\lambda x\del_{-\lambda}a=\int xa-\lambda\int \del_{-\lambda}a=\int xa \\
\int x\frac{\del a}{\del x} &=\int \frac{\del }{\del x}xa-\int a=-\int a\\
\int ta &=\int at-\lambda x\frac{\del a}{\del x }=\int at+\lambda\int a\\
\int  (\del_{\pm\lambda} a)t&=\pm\int (a(x,t\pm\lambda)-a(x,t)){t\over\lambda}=\pm\lambda^{-1}\int (a(x,t\pm\lambda)(t\pm \lambda)-a(t)t)-\int a(t\pm\lambda)=-\int a.
\end{align*}

Then for the $\alpha=1$,
\begin{align*}
\int a\X&-\X^*a+X[\extd x,a]=\int a(u_1+c_{11}x+c_{12}t)-(\overline{u_1}+\overline{c_{11}}x+\overline{c_{12}}t)a+\lambda \frac{\partial}{\partial x}a(x,t+\lambda)\Theta\\
&=\int (u_1-\overline{u_1})a+\int (c_{11}-\overline{c_{11}})ax+c_{12}at-\overline{c_{12}}ta+\lambda(\frac{\partial}{\partial x}a(x,t+\lambda))(w+c_1 x+ c_2 t)\\
&=\int (u_1-\overline{u_1})a+\int (c_{11}-\overline{c_{11}})xa+\int (c_{12}-\overline{c_{12}})at-\overline{c_{12}}\lambda \int a-c_1\lambda\int a
\end{align*}
and vanishing of this for all $a$ leads to 
\[u_1-\overline{u_1}=\lambda(c_{12}+c_1),\quad c_{11}=\overline{c_{11}},\quad c_{12}=\overline{c_{12}}.\]
For $\alpha=2$, 
\begin{align*}
\int a\Y-\Y^*a+X[\extd t,a]&=\int a(u_2+c_{21}x+c_{22}t)-(\overline{u_2}+\overline{c_{21}}x+\overline{c_{22}}t)a\\
&\quad -\lambda\left({\del a\over \del x}\X+(\del_{-\lambda} a) \Y\right)-\frac{\lambda}{2}\left(\lambda{\del^2 a\over\del x^2}+(\del_{\lambda}+\del_{-\lambda})a+\lambda\frac{\del^2}{\del x^2}\del_\lambda a\right)\Theta\\
&=\int (u_2-\overline{u_2})a+\int (c_{21}-\overline{c_{21}})xa+c_{22}at-\overline{c_{22}}ta\\
&\quad -\lambda \left(x{\del a\over \del x}c_{11}+(\del_{-\lambda} a) c_{22} t\right)-\frac{\lambda}{2}\left(\lambda c_1x{\del^2 a\over\del x^2}+c_2(\del_{\lambda}+\del_{-\lambda})at\right)\\
&=\int (u_2-\overline{u_2})a+\int (c_{21}-\overline{c_{21}})xa+\int (c_{22}-\overline{c_{22}})at\\
&\quad +c_{11}\lambda\int a+\lambda (c_{22}-\overline{c_{22}})\int a+\lambda c_{2}\int a
\end{align*}
and vanishing of this leads to
\[u_2-\overline{u_2}=-\lambda(c_{11}+c_2),\quad c_{21}=\overline{c_{21}},\quad c_{22}=\overline{c_{22}}.\]
For $\alpha=3$, 
\begin{align*}
\int a\Theta-\Theta^*a&+X[\theta',a]=\int a(w+c_{1}x+c_{2}t)-(\overline{w}+\overline{c_{1}}x+\overline{c_{2}}t)a+\lambda (\del_\lambda a)\Theta\\
&=\int (w-\overline{w})a+\int (c_{1}-\overline{c_{1}})ax+\int (c_{2}-\overline{c_{2}})at-\overline{c_{2}}\lambda \int a-c_2\lambda\int a
\end{align*}
and vanishing of this leads to
\[w-\overline{w}=2\lambda c_2,\quad c_{1}=\overline{c_{1}},\quad c_{2}=\overline{c_{2}}.\]
This completes the derivation of the reality requirements as stated.  \endproof

We can now solve the quantum geodesic auxiliary and velocity equations (\ref{bicvel1})--(\ref{bicvel2}).

\begin{proposition}\label{solX}  All possible quantum geodesic velocity fields, i.e. obeying (\ref{bicvel1})--(\ref{bicvel2}),   that are constant plus linear in $x,y$ are in the following table:
\[ \begin{array}{c||c|c|c|c|c}& \X & \Y & \Theta & \kappa&  {\rm Timelike\ or\ null}\\
\hline\hline
A\phantom{\Big|}& u_1 & u_2 & w & 0& |u_2|\ge |u_1|\\ \hline 
B\phantom{\Big|}& u_1 & a+b(x - u_1s) &  0& 0 & |a+bx|\ge |u_1|\\ \hline
C\phantom{\Big|}& a+b{\lambda\over 2}+b (t-v x)&  v \X& 0& -b v(1+v)& |v|\ge 1\\ \hline
D\phantom{\Big|}&a+b{\lambda\over 2}+b(t-u_2 s) &u_2 &-\Y & 0& |u_2|\ge |a+bt |\\ \hline
E \phantom{\Big|}& u_1 & \frac{t+d -b(x-u_1s)}{s-a}& 0& \frac{1}{s-a}& |t+d -b x|\ge  |a u_1|  \\ \hline
F\phantom{\Big|}& u_1& \frac{t}{s-a} & -2 \Y & \frac{1}{s-a}& |t|\ge |a u_1| \\ \hline
G \phantom{\Big|}& u_1& \frac{b+\lambda+t}{s-a}& -2\Y & \frac{1}{s-a}& |b+t|\ge |a u_1|\\ \hline
H\phantom{\Big|}& \frac{x}{s-a} & \frac{b-\frac{\lambda}{2}+t}{s-a}+\frac{bx}{(s-a)^2}& 0& \frac{2}{s-a}& |b+t- { b x\over a}|\ge |x|
\end{array}\]
where  $u_1,u_2,v, w,a,b,d$ are now denote some  real constant parameters. The solutions all have classical limits as $\lambda\to 0$ and we show the condition for that to be timelike or null. 
\end{proposition}
\proof  For geodesic velocity fields of the constant plus linear in $x,y$ form, according to (\ref{divnablaf}) we require ${\rm div}_\nabla(X_s)=\Tr(C)1$, and if we take $\kappa\in \C.1$ then the first half of the unitarity conditions  (\ref{kappaalpha}) forces us to 
 $\kappa=\frac{1}{2}\Tr(C)$. We use that $\del_3\Theta=0$ under our linear assumption. 
 
 Since $\kappa\in \C.1$, it does not enter the velocity equation. We next work out the commutators
 \[ [\extd x, X^i]=\lambda c_{i1}\theta',\quad [\extd t, X^i]=-\lambda \extd x c_{i1}-\lambda(\theta'+\extd t)c_{i2}\]
 \[[\extd x,\Theta]=\lambda c_1\theta',\quad [\extd t,\Theta]=-\lambda\extd x c_1-\lambda (\theta'+\extd t)c_2\]
 where $X^1=\X, X^2=\Y$ is a shorthand and we leave understood the possible $s$-dependence. Now putting in the form of $X^i$ and matching coefficients of $1,x,t$, the auxiliary equations (\ref{bicvel1}) become the 
 9 equations
\begin{align}
 c_{21} w + c_{11} u^1 +  c_{12} (w + u^2 ) =0,\quad  (c_{11} c_{22} - c_{12} c_{21}) = (c_{21} + c_{12}) c_1+ c_{11}^2 + c_{12} c_{21},\label{9equaitons1}\\
 \quad  (c_{12} + c_{21}) c_2   + c_{12} (c_{11} + c_{22}) =     0,\quad  (c_1  - c_{21}) w = 0,\quad c_{11} c_2   = c_1 ^2,\quad  (c_1  - c_{21}) c_2=0\label{9equaitons2}\\
c_1  u^1 + c_2 u^2 + (c_2   + c_{22}) w = 0,\quad ( 2 c_{21} + c_1 ) c_2   + c_1  c_{11} = 0,\quad c_2   (c_2   + 2 c_{22}) + c_1  c_{12} = 0\label{9equaitons3}
\end{align}
 Now we look at the velocity equation (\ref{bicvel2}). Since
 \[ \extd X^i=c_{i1}\extd x+ c_{i2}\extd t,\quad \extd\Theta=c_1 \extd x+ c_2\extd t\]
 and $\kappa$ was assumed central, we obtain 9 equations. 6 of them are the same as (\ref{cueqn})
 for the classical case. In addition we have
 \begin{align}
 {\del w\over\del s}+ c_i u^i=0,\quad {\del c_j\over\del s} + c_i c_{ij}=0\label{2equations}
 \end{align}
 (sum over $i$).
 
 Then (\ref{cueqn}),(\ref{newunitarity}),(\ref{9equaitons1}),(\ref{9equaitons2}),(\ref{9equaitons3}),(\ref{2equations}) are the full set of equations we need to solve. First of all, considering (\ref{newunitarity}), the last two equations of (\ref{9equaitons2}) and second equation of (\ref{9equaitons3}) give 
 \begin{align}
 c_1=0
 \end{align}
 From the first equation of (\ref{2equations}), we have\[\frac{\del (w-\overline w)}{\del s}+c_2 (u^2-\overline {u^2})=0\]
 Using (\ref{newunitarity}) it becomes\[2\frac{\del c_2}{\del s}-c^2_2-c_2 c_{11}=0\]
 Considering the third equation of (\ref{9equaitons2}), it gives $c_2=0$ or $c_2=-\frac{2}{s-a}$. If $c_2=-\frac{2}{s-a}$, the remaining equations give the last two solutions stated. If $c_2=0$, the equations give the second, third, fourth and fifth solutions stated. 
 
All solutions have a classical limit as $\lambda\to 0$ because our velocity equations from (\ref{bicvel2}) included the classical ones, and the final column shows which of these limits are timelike or null. We already explained in Section~\ref{secmink} how to do this. Namely, we just need $(\X(0),\Y(0))$ as $s=0$ to be timelike or null. Finally, we present the solutions in terms of fresh real parameters renamed bearing in mind their role in the classical limit. \endproof

Compared with the completely classical flat case of Section~\ref{secflat}, we see that the $X_s$ that arise are special cases. For example, they do not include the whole family (\ref{excu}) after replacing the geodesic time $t$ there by $s$.   Thus  {\em not every classical geodesic flow can be quantised} to this quantum spacetime in the strict sense of both the velocity equation and the auxiliary one.   

We next consider the associated quantum geodesic amplitude flows in H case in Proposition~\ref{solX}. Then (\ref{bicflow}) becomes
\begin{align*}
{\del \psi\over\del s}&=-{\del \psi\over\del x} \X - (\del_{-\lambda}\psi) \Y  - \psi \kappa_s\\
&=-{\del \psi\over\del x}\frac{x}{s-a}-(\del_{-\lambda}\psi)\left(\frac{b-\frac{\lambda}{2}+t}{s-a}+\frac{bx}{(s-a)^2}\right)-\frac{2}{s-a}\psi\\
&=\frac{-1}{s-a}\left(x{\del \psi\over\del x}+\left[{\del \psi\over\del x},x\right]\right)-(\del_{-\lambda}\psi)\frac{b-\frac{\lambda}{2}+t}{s-a}-\frac{b}{(s-a)^2}\left(x\del_{-\lambda}\psi+[\del_{-\lambda}\psi,x]\right)-\frac{2}{s-a}\psi\\
&=\frac{-x}{s-a}{\del \psi\over\del x}+\lambda\frac{x}{s-a}\del_{-\lambda}{\del \psi\over\del x}-(\del_{-\lambda}\psi)\frac{b-\frac{\lambda}{2}+t}{s-a}-\frac{bx}{(s-a)^2}\del_{-\lambda}\psi+\lambda\frac{bx}{(s-a)^2}\del_{-\lambda}\del_{-\lambda}\psi-\frac{2}{s-a}\psi\\
&=\frac{-x}{s-a}{\del \psi\over\del x}-\frac{x}{s-a}\sum_{n=1}^{\infty}\frac{\del^{n+1}\psi(x,t)}{\del x\del t^n}\frac{(-\lambda)^n}{n!}-\sum_{n=1}^{\infty}\frac{\del^n\psi(x,t)}{\del t^n}\frac{(-\lambda)^{n-1}}{n!}\frac{b-\frac{\lambda}{2}+t}{s-a}\\
&-\frac{bx}{(s-a)^2}\sum_{n=1}^{\infty}\frac{\del^n\psi(x,t)}{\del t^n}\frac{(-\lambda)^{n-1}}{n!}-\frac{bx}{(s-a)^2}\sum_{m,n=1}^{\infty}\frac{(-\lambda)^{n+m-1}}{n!m!}\frac{\del^{m+n}\psi}{\del t^{m+n}}-\frac{2}{s-a}\psi
\end{align*}
At classical level where $\lambda$ goes to $0$, it reduces to
\begin{align}\label{solution1}
{\del \psi^c\over\del s}=\frac{-x}{s-a}{\del\psi^c\over\del x}-{\del\psi^c\over\del t}\frac{b+t}{s-a}-\frac{bx}{(s-a)^2}{\del\psi^c\over\del t}-\frac{2}{s-a}\psi^c
\end{align}
with solution
\begin{align*}
\psi^c=\left(\frac{a}{s-a}\right)^2\psi^c_0\left(u,v\right)
\end{align*} 
where

\[(u,v)=\left(\frac{a x}{a-s},\frac{at}{a-s}-sb\left(\frac{1}{s-a}+\frac{x}{(s-a)^2}\right)\right)=\phi_s(x,t)\]
\[ \phi^{-1}_s(x,t)=\left(-\frac{x(s-a)}{a},t\frac{a-s}{a}+\frac{sbx}{a^2}-\frac{sb}{a}\right)=(x,t)+s(\X(0),\Y(0)). \]
As we saw already in Section~\ref{secflat} this represents linear motion as $s$ increases, eg an initial bump function $\psi^c_0$ sees this bump moving with a constant vector.

Setting $\psi=\psi^c+\lambda \psi^1$, (\ref{bicflow}) becomes
\begin{align*}
{\del(\psi^c+\lambda \psi^1) \over\del s}&=-\frac{x}{s-a}{\del (\psi^c+\lambda \psi^1)\over\del x}-\frac{x}{s-a}\sum_{n=1}^{\infty}\frac{\del^{n+1}(\psi^c+\lambda \psi^1)}{\del x\del t^n}\frac{(-\lambda)^n}{n!}\\
&-\sum_{n=1}^{\infty}\frac{\del^n(\psi^c+\lambda \psi^1)}{\del t^n}\frac{(-\lambda)^{n-1}}{n!}\frac{b-\frac{\lambda}{2}+t}{s-a}-\frac{bx}{(s-a)^2}\sum_{n=1}^{\infty}\frac{\del^n(\psi^c+\lambda \psi^1)}{\del t^n}\frac{(-\lambda)^{n-1}}{n!}\\
&-\frac{bx}{(s-a)^2}\sum_{m,n=1}^{\infty}\frac{\del^{m+n}(\psi^c+\lambda\psi^1)}{\del t^{m+n}}\frac{(-\lambda)^{n+m-1}}{n!m!}-\frac{2}{s-a}(\psi^c+\lambda \psi^1)
\end{align*}
At the level of the order $\lambda$, this equation becomes
\begin{align*}
{\del\psi^c \over\del s}+\lambda{\del\psi^1 \over\del s}&=-\frac{x}{s-a}{\del \psi^c\over\del x}-\lambda\frac{x}{s-a}{\del \psi^1\over\del x}+\lambda\frac{x}{s-a}\frac{\del^2\psi^c}{\del x\del t}-\frac{\del\psi^c}{\del t}\frac{b+t}{s-a}+\frac{\lambda}{2(s-a)}\frac{\del\psi^c}{\del t}\\
&\quad-\lambda\frac{\del\psi^1}{\del t}\frac{b+t}{s-a}+\frac{\lambda}{2}\frac{\del^2\psi^c}{\del t^2}\frac{b+t}{s-a}-\frac{bx}{(s-a)^2}\left(\frac{\del\psi^c}{\del t}+\lambda\frac{\del\psi^1}{\del t}-\frac{\lambda}{2}\frac{\del^2\psi^c}{\del t^2}\right)\\
&\quad+\frac{bx}{(s-a)^2}\frac{\del^2\psi^c}{\del t^2}\lambda-\frac{2}{s-a}(\psi^c+\lambda \psi^1)
\end{align*}
Using(\ref{solution1}) we get
\begin{align*}
{\del\psi^1 \over\del s}&=-\frac{x}{s-a}{\del \psi^1\over\del x}+\frac{x}{s-a}\frac{\del^2\psi^c}{\del x\del t}+\frac{1}{2(s-a)}\frac{\del\psi^c}{\del t}-\frac{\del\psi^1}{\del t}\frac{b+t}{s-a}+\frac{1}{2}\frac{\del^2\psi^c}{\del t^2}\frac{b+t}{s-a}\\
&\quad -\frac{bx}{(s-a)^2}\frac{\del\psi^1}{\del t}+\frac{3bx}{2(s-a)^2}\frac{\del^2\psi^c}{\del t^2}-\frac{2}{s-a} \psi^1
\end{align*}
The solution is
\begin{align}
\psi^1(x,t)=\left(\frac{a}{s-a}\right)^2\psi^1_0(u,v)+\frac{1}{2}\left(\frac{a}{s-a}\right)^3\left(\frac{\del\psi^c_0(u,v)}{\del v}+\left(b+\frac{b(s+a)}{a(s-a)}u+v\right)\frac{\del^2\psi^c_0(u,v)}{\del v^2}+2u\frac{\del^2\psi^c_0(u,v)}{\del u\del v}\right)
\end{align}

In the case where $\psi^1_0=0$ and initially $\psi^0_0$ is a Gaussian form $e^{-\frac{x^2}{\beta}}e^{-\frac{t^2}{\beta}}$, we have
\begin{align*}
\psi^1(x,t)&=\frac{1}{(s-a)^2\beta^2}\Big(2v[(bsuv-a^2(2u^2+v(b+v)+a(2su^2+buv+sv(b+v))]\\
&\qquad+[ab(a-s)-b(a+s)u+2a(a-s)v]\beta\Big)\psi^c(x,t)
\end{align*}
where $\psi^c(x,t)=\frac{a^2}{(s-a)^2}e^{-\frac{u^2}{\beta}}e^{-\frac{v^2}{\beta}}$.

In the case where $\psi^1_0=0$ and initially $\psi^0_0$ is a plane wave $e^{- \imath k x}e^{\imath \omega t}$, we have
\begin{align*}
\psi^1(x,t)&=\frac{ w}{2(s-a)^2}\Big(\imath a(s-a)+ab(a-s)\omega+u(2ak(s-a)-b(s+a)\omega)+v\omega a(a-s)\Big)\psi^c(x,t)
\end{align*}
where $\psi^c(x,t)=\frac{a^2}{(s-a)^2}e^{- \imath k u}e^{\imath \omega v}$. Thus, we have
\begin{align*}
\psi_s&(x,t)=\psi^c(x,t)+\lambda\psi^1(x,t)\\
&=\left(1+\lambda \frac{ w}{2(s-a)^2}\Big(\imath a(s-a)+ab(a-s)\omega+u(2ak(s-a)-b(s+a)\omega)+v\omega a(a-s)\Big)\right)\frac{a^2}{(s-a)^2}e^{- \imath k u}e^{\imath \omega v}\\
&\approx \frac{a^2}{(s-a)^2}e^{\frac{\lambda \omega}{2(s-a)^2}(\imath a(s-a)+ab(a-s)\omega)}e^{\frac{\lambda \omega}{2(s-a)^2}(u(2ak(s-a)-b(s+a)\omega)+v\omega a(a-s))}e^{- \imath k u}e^{\imath \omega v}\\
&=\frac{a^2}{(s-a)^2}e^{\frac{\lambda_p a\omega}{2(a-s)}(1+\imath\omega b)}e^{- \imath uk[1+\frac{\lambda_p  \omega}{(s-a)^2 k}(b(s+a)\omega-2ak(s-a))]}e^{\imath  v\omega(1-\frac{\lambda_p a\omega}{2(s-a)} )}\\
&=\frac{a^2}{(s-a)^2}e^{\frac{\lambda_p a\omega}{2(a-s)}(1+\imath\omega b)}e^{- \imath uk'}e^{\imath  v\omega'}\\
\end{align*}
working up to order $\lambda_p/(s-a)$. Here the effective momentum and frequency at this order are 
\[ k'=k\left(1+\frac{\lambda_p  \omega}{(s-a)^2 k}(b(s+a)\omega-2ak(s-a))\right),\quad \omega'=\omega\left(1-\frac{\lambda_p a\omega}{2(s-a)} \right).\]
These are changes in the normal ordered form, but they indicate an order $\lambda_p/(s-a)$ correction also in more physical values determined, for example, by an analysis of expectation values as we saw for the constant case. We also see that the factor at the front of $\psi$ is not a pure phase, which is a new phenomenon compared to our previous cases. Also, the corrections blow up at the singularity as $s\to a$. This means that, at least within perturbation theory, we cannot see what happens to $\psi$ as we approach such points due to quantum gravity effects if $\lambda_p$ is of order Planck scale.

\section{Quantum geodesics on the noncommutative torus} \label{sectorus}

This is not a quantum spacetime but included by way of contrast as a curved example of Section~\ref{secpara}. The noncommutative torus is the Weyl form of the canonical commutation relations algebra with unitary generators $u,v$ and relations $vu=e^{\imath\theta}uv$. We equip it with the 2D differential  calculus where $\Omega^1$ has a central basis
\begin{align*}
e^1=u^{-1}\extd u ,\quad e^2=v^{-1}\extd v
\end{align*}
and a $*$-structure with $u^*=u^{-1}, v^*=v^{-1}$ and $e^i{}^*=-e^i$. We then take a generic (symmetric) quantum metric and the moduli of WQLCs for it found in  \cite[Ex.~ 8.16]{BegMa},
\begin{align*}
g&=c_1e^1\tens e^1+c_2e^2\tens e^2+c_3(e^1\tens e^2+e^2\tens e^1)\\
\nabla e^i&=h^i{}_1 e^1\tens e^1+h^i{}_2 e^2\tens e^2+h^i{}_3(e^1\tens e^2+e^2\tens e^1)
\end{align*}
where $i=1, 2$, $c_i$ are real for the `reality' of the metric  and, for a given metric, the $h^i{}_j$ need to obey
\begin{align}\label{c_{12}3}
c_2 h^2{}_1=c_1 h^1{}_3+c_3(h^2{}_3-h^1{}_1),\quad c_2 h^2{}_3=c_1 h^1{}_2+c_3(h^2{}_2-h^1{}_3) 
\end{align}
and be imaginary for the `reality' properties of a connection with respect to $*$. The curvature $R_\nabla$ of the above connection is also computed in \cite{BegMa}, as
\[R_\nabla(e^i)=\rho^i_j e^1\wedge e^2\tens e^j;\quad \rho=S\begin{pmatrix} c_3&c_2\\-c_1&-c_3\end{pmatrix},\quad  S=\begin{cases}\frac{h^1_2 h^2_1-h^1_3 h^2_3}{c_3} & c_3\ne0\\ 
\frac{c_1 h^1_3(h^1_3-h^2_2)+c_2 h^2_3(h^2_3-h^1_1)}{c_1c_2} & c_3=0\end{cases}\]
The case $h^i{}_j=0$ is the trivial flat QLC but $\nabla$ for the quantum `geodesic' flow (now more properly
called an autoparallel flow) can be any other connection and here we choose a family with a slightly weaker form
of metric compatibility in order to exhibit curvature.   

Next, we analyse quantum geodesics. For brevity, we let
\begin{align*}
U=X_t(e^1)=X^1_t,\quad V=X_t(e^2)=X^2_t
\end{align*}
as elements of $A$. Then the auxiliary equation (\ref{veleq1}) becomes
\begin{align}\label{uv0}
[U,V]=0
\end{align}
and the velocity equations (\ref{veleq2}) become
\begin{align}
\dot U+[U,\kappa_t]+u\frac{\partial U}{\partial u}U+v\frac{\partial U}{\partial v}V+h^1{}_1 U^2+h^1{}_2 V^2+h^1{}_3 UV=0\label{U}\\
\dot V+[V,\kappa_t]+u\frac{\partial V}{\partial u}U+v\frac{\partial V}{\partial v}V+h^2{}_1 U^2+h^2{}_2 V^2+h^2{}_3 UV=0\label{V}
\end{align}

Next, we define $\int$ by \[\int u^m v^n=\delta_{m0}\delta_{n0}\] for any $m,n$ (corresponding classically to integration over a torus with its usual Haar measure) and require $\Gamma^\beta_{\beta\alpha}=0$ in Section~\ref{secpara}, which means\[h^1_1+h^2_3=0,\quad h^2_2+h^1_3=0,\] so that the divergence condition (\ref{divcondGamma})  holds. Because $e^i$ are an anti-self adjoint central basis, the second half of the unitarity condition (\ref{unitarity}) reduces to 
\begin{equation}\label{reqX}
X_t^\alpha=-X_t^\alpha{}^*
\end{equation}
From the first half of (\ref{unitarity}), we then have $\kappa_t=\frac{1}{2}\del_\alpha X^\alpha$. 

The final step is to solve the amplitude flow equation $\nabla_E\psi=0$  for $\psi(t)\in A$, which becomes
\begin{align}\label{solve}
 \dot\psi=-u\frac{\partial \psi}{\partial u}U-\frac{\partial \psi}{\partial v}vV-\psi \kappa_t,
\end{align}
where at each time $\psi(t)(u,v)$ is normal ordered with $v$ to the right. 

For simplicity, we illustrate some specific results in the standard metric case $c_1=c_2\ne 0, c_3=0$. Then (\ref{c_{12}3}) becomes
\[h^2{}_1=h^1{}_3,\quad h^1{}_2=h^2{}_3,\]
which one can check from the above is still curved for suitable $h^i{}_j$.  We also restrict to the easy case where $U,V,\kappa\in \C.1$ in $A$ are constant on the torus (they could be time-varying). Then (\ref{reqX}) holds and $\kappa=0$, while (\ref{U}),(\ref{V}) reduce to
\begin{align*}
\dot U+h^1{}_1 U^2-h^1{}_1 V^2-h^2{}_2 UV=0\\
\dot V-h^2{}_2 U^2+h^2{}_2 V^2-h^1{}_1 UV=0, 
\end{align*}
while a compatible $*$ on vector fields requires $U,V$ imaginary also. Hence we replace 
\[ h^a{}_b=\imath H^a{}_b,\quad U=\imath X,\quad V=\imath Y\]
so that the above equations become
\begin{align*}
\dot X-H^1{}_1 X^2+H^1{}_1 Y^2+H^2{}_2 XY=0\\
\dot Y+H^2{}_2 X^2-H^2{}_2 Y^2+H^1{}_1 XY=0,
\end{align*}
where now $X(t),Y(t)$ are real solutions of these equations with $H^i{}_j$ real parameters.   We can solve these numerically and then proceed to the amplitude flow. 

\begin{proposition} \label{solve1}
 The amplitude flow solving (\ref{solve}) for $U,V,\kappa\in \C.1$ is 
\[ \psi_t (u,v)=\psi_0 (u e^{-i\int_0^t X dt},v e^{-i\int_0^t Y dt})\]
for an initial normal ordered function $\psi_0$. Moreover, at geodesic time $t$, 
\[ \<u^mv^n\>_t=\<u^mv^m\>_0 e^{\imath m \int_0^t X dt} e^{\imath n\int_0^t Y dt}.\]
\end{proposition}
\proof  We set $\psi (u,v)=\Sigma \psi_{m,n}u^m v^n$, then (\ref{solve}) becomes
\begin{align*}
\Sigma \dot\psi_{mn}u^m v^n&=\Sigma-umu^{m-1}v^n\psi_{mn}U-\psi_{mn}u^m nv^{n-1}vV\\
&=\Sigma\psi_{mn}(-m)u^m v^nU-n\psi_{mn}u^mv^nV=\Sigma\psi_{mn}u^mv^n(-mU-nV).
\end{align*}
Hence
\[\dot\psi_{mn}=\psi_{mn}(-mU-nV);\quad \psi_{mn}=e^{-\int dt(mU+nV)}\psi(0)_{mn}\]
and
\begin{align*}
\psi(u,v)&=\Sigma e^{-\int dt(mU+nV)}\psi(0)_{mn}u^m v^n=\Sigma\psi(0)_{mn}(e^{-\int dtU}u)^m (e^{-\int dtV}v)^n\\
&=\psi_0(ue^{-\int dtU},ve^{-\int dtV})=\psi_0(ue^{-i\int dtX},ve^{-i\int dtY}).
\end{align*}

The expectation values stated in the last part are an easy calculation similarly writing $\psi_0(u,v)=\sum_{p,q}\psi_{p,q}u^pv^q$ and computing $\<u^mv^n\>=\int \psi_t^* u^m v^n \psi_t/ \int \psi_t^* \psi_t$. The denominator here does not depend on $t$ as we also know from Lemma~\ref{lemuni}.  \endproof

\begin{figure}
\[ \includegraphics[scale=0.8]{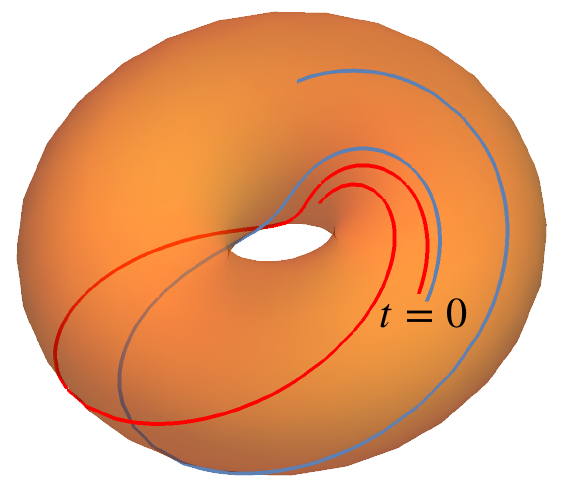}\]
\caption{Numerical solutions for a geodesic flow starting at $X(0)=Y(0)=1$ and a nearby one at $X(0)=1.1$. Here $H^1{}_1=0.02$ and $H^2{}_2=0.04$. \label{p2}}
\end{figure}

In Figure~\ref{p2}, we plot a couple of geodesics flows with $\theta(t)=-\int_0^t X \extd t$, $\phi(t)=-\int_0^t Y \extd t$ depicted as angles on a torus. The geodesic flow is multiplication of $(u,v)$ in the algebra by $(e^{\imath\theta(t)},e^{\imath\phi(t)})$ depicted here as an element of a classical torus. Notice the large amount of geodesic deviation as a signal of curvature. 
In the flat case, where $h^i{}_j=0$, we would just get $
\dot X=\dot Y=0$  and hence
\begin{align*}
\psi_t (u,v)=\psi_0 (u e^{-i t X},v e^{-i t Y}),
\end{align*}
which is like  uniform geodesic flow on a classical torus even though, here, $u,v$ are normal ordered and noncommutative. 
 
\section{Concluding remarks}\label{secrem}

Usually in GR, we use the full power of Riemannian geometry in order to treat curved spacetimes in a coordinate-invariant manner. In this paper we have seen that the same geometric approach, now quantum Riemannian geometry,  is equally useful for flat but noncommutative spacetimes where it keeps our constructions abstractly defined and thereby coordinate invariant. The effective momentum space  here is (in some sense) curved  but from the position point of view there are now complexities from the noncommutation relations so that one might equally wonder how much constructions depend on the choices of $x_i,t$. Our approach depends only on the algebra $A$, $\Omega^1$ as a bimodule, etc., and one is free to choose any generators of the algebra and any basis for the 1-forms here. For example, in \cite{AmeMa},  there was an ad-hoc assumption that the physics would correspond to the noncommutative algebra when plane waves were normal ordered,  in order to make the variable speed of light prediction there.  We have replaced that by the quantum geodesics formalism and found that for plane waves there {\em is} an order $\lambda_p\omega$ effect but it is to the effective geodesic velocity not to the plane wave itself. This, however, should be what is relevant to the time-of-flight experiment for cosmological $\gamma$-ray bursts analysed in \cite{AmeMa}. We also did it for Gaussians and found an order $\lambda_p^2/\beta$ effect. This calculation implies that a point particle can not, physically, be modelled as the limit of increasingly sharp Gaussians due to quantum gravity corrections in so far as these are expressed in the choice of quantum spacetime. We did make use of normal ordering to specify formulae, for example to compute partial derivatives with respect to our basis or to define  integration over the spacetime but this was for convenience and the constructions themselves are invariantly defined. Also, for the initial plane-wave and initial Gaussian, we took the normal ordered $e^{-\imath k x}e^{\imath \omega t}$ and $e^{-{x^2\over\beta}}e^{-{t^2\over\beta}}$ respectively, but the theory allows input of any initial $\psi$ for the geodesic evolution. 

Note that geometric or coordinate invariance of our constructions should not be confused with physical symmetries such as Lorentz invariance. Clearly our constructions are not classically Lorentz invariant and take place in a particular inertial frame $x,t$. There is  a quantum Poincar\'e group symmetry under which the differential calculus and flat connection that we used are invariant, but exactly how this manifests in the quantum geodesic formalism remains to be explored. Also, although we did our calculations for $x,t$,  they work the same way for $x^i$ as these commute amongst themselves and each behave the same way with respect to $t$ in the algebra. 
 
Many of the unusual features of the paper are present even classically and come from the unusual point of view on geodesic flows, where we first find the velocity fields $X$ and only retrospectively couple them to the amplitudes $\psi$. Amplitudes themselves are spread over spacetime not space, and their interpretation as probability is with respect to the geodesic time $s$. There are several conceptual issues with this but {\em mathematically} in the classical case this is equivalent to working with collections of ordinary geodesics. Therefore we studied the classical case in some depth also, in Section~\ref{secflat}. One of the effects here is that the velocity field $X$ could have points of infinite divergence as the system evolves, i.e. as $s\to a$. Physically this corresponds to some kind of source or sink in the sense that ${\rm div}(X)$ also blows up, but the phenomenon should be studied further even in the classical case. In the paper we limited concrete calculations  to $X$ either constant or constant plus linear in the spacetime variables, but clearly more general classes of classical $X$ and their physical interpretation deserve to be investigated. 

At the quantum level on $\lambda$-Minkowski spacetime we found that the geodesic velocity equations are more restrictive with fewer allowed $X$ in the class that we looked at. To a large extent, this can be attributed to the auxiliary equation (\ref{auxeqn}) and in \cite{BegMa:ric} it is shown that, while lacking the elegant simplicity of $\doublenabla(\sigma_E)=0$, this part of a quantum geodesic can be relaxed to a weaker condition. We have not analysed exactly which classical geodesics are then quantisable. We did, however, find interesting solutions for $X$  even of the stricter original version, and then found that the perturbative corrections to the associated amplitude flows for $\psi$ blow up at the points of singularity {\em faster} than the classical $\psi$, i.e. quantum gravity corrections will dominate as the geodesic time $s\to a$ at such singular solutions. This phenomenon again deserves to be investigated further. 

In short, while the quantum geodesic formalism is new and so far little-explored, our test-drive of it on $\lambda$-Minkowski spacetime, as well as the noncommutative torus case put in for contrast, shows that it is capable of interesting answers and deserves to be explored in other models, both classically on a general (pseudo) Riemannian manifold due to the new point of view and on quantum space and quantum spacetime models, including finite ones. The case the equiliateral triangle graph $\Z_3$ was in the original work \cite{Beg:geo} while some further models including quantum geodesics on the fuzzy sphere recently appeared in \cite{BegMa:ric}.

\end{document}